\documentclass[3p,final,twocolumn]{elsarticle}

\usepackage[hidelinks]{hyperref}
\usepackage{url}

\providecommand{\bibcommenthead}{}%
\providecommand{\burl}[1]{\url{#1}}%

\makeatletter\if@twocolumn\PassOptionsToPackage{switch}{lineno}\else\fi\makeatother

\usepackage{tabulary,xcolor}
\usepackage{amsfonts,amsmath,amssymb}
\usepackage[T1]{fontenc}
\makeatletter
\let\save@ps@pprintTitle\ps@pprintTitle

\def\ps@pprintTitle{\save@ps@pprintTitle\gdef\@oddfoot{\footnotesize\itshape \null\hfill\raisebox{-40pt}{\thepage}}}

\def\hlinewd#1{%
  \noalign{\ifnum0=`}\fi\hrule \@height #1%
  \futurelet\reserved@a\@xhline}

\makeatother

\usepackage{ifluatex}
\ifluatex
\usepackage{fontspec}
\defaultfontfeatures{Ligatures=TeX}
\usepackage[]{unicode-math}
\unimathsetup{math-style=TeX}
\else 
\usepackage[utf8]{inputenc}
\fi 
\ifluatex\else\usepackage{stmaryrd}\fi
\usepackage[font=small]{caption}  

\usepackage[english]{babel}
\usepackage[autostyle=true]{csquotes} 

\usepackage{url,multirow,morefloats,floatflt,cancel,tfrupee}
\makeatletter

\AtBeginDocument{\@ifpackageloaded{textcomp}{}{\usepackage{textcomp}}}
\makeatother
\usepackage{colortbl}
\usepackage{xcolor}
\usepackage{pifont}
\usepackage[nointegrals]{wasysym}
\makeatletter

\def\mcWidth#1{\csname TY@F#1\endcsname+\tabcolsep}

\def\cAlignHack{\rightskip\@flushglue\leftskip\@flushglue\parindent\z@\parfillskip\z@skip}
\def\rAlignHack{\rightskip\z@skip\leftskip\@flushglue \parindent\z@\parfillskip\z@skip}

\@ifundefined{etal}{}{}

\usepackage{ifxetex}
\ifxetex\else\if@twocolumn\@ifpackageloaded{stfloats}{}{\usepackage{dblfloatfix}}\fi\fi

\AtBeginDocument{
\expandafter\ifx\csname eqalign\endcsname\relax
\def\eqalign#1{\null\vcenter{\def\\{\cr}\openup\jot\m@th
  \ialign{\strut$\displaystyle{##}$\hfil&$\displaystyle{{}##}$\hfil
      \crcr#1\crcr}}\,}
\fi
}

\AtBeginDocument{%
  \@ifpackageloaded{endfloat}%
   {\renewcommand\efloat@iwrite[1]{\immediate\expandafter\protected@write\csname efloat@post#1\endcsname{}}}{\newif\ifefloat@tables}%
}%

\def\BreakURLText#1{\@tfor\brk@tempa:=#1\do{\brk@tempa\hskip0pt}}
\let\lt=<
\let\gt=>
\def\processVert{\ifmmode|\else\textbar\fi}

\@ifundefined{subparagraph}{
\def\subparagraph{\@startsection{paragraph}{5}{2\parindent}{0ex plus 0.1ex minus 0.1ex}%
{0ex}{\normalfont\small\itshape}}%
}{}

\newcommand\role[1]{\unskip}
\newcommand\aucollab[1]{\unskip}
  
\@ifundefined{tsGraphicsScaleX}{\gdef\tsGraphicsScaleX{1}}{}
\@ifundefined{tsGraphicsScaleY}{\gdef\tsGraphicsScaleY{.9}}{}

\def\checkGraphicsWidth{\ifdim\Gin@nat@width>\linewidth
	\tsGraphicsScaleX\linewidth\else\Gin@nat@width\fi}

\def\checkGraphicsHeight{\ifdim\Gin@nat@height>.9\textheight
	\tsGraphicsScaleY\textheight\else\Gin@nat@height\fi}

\def\fixFloatSize#1{}
\let\ts@includegraphics\includegraphics

\def\inlinegraphic[#1]#2{{\edef\@tempa{#1}\edef\baseline@shift{\ifx\@tempa\@empty0\else#1\fi}\edef\tempZ{\the\numexpr(\numexpr(\baseline@shift*\f@size/100))}\protect\raisebox{\tempZ pt}{\ts@includegraphics{#2}}}}

\AtBeginDocument{\def\includegraphics{\@ifnextchar[{\ts@includegraphics}{\ts@includegraphics[width=\checkGraphicsWidth,height=\checkGraphicsHeight,keepaspectratio]}}}

\usepackage[font=small,labelfont=bf]{caption}

\usepackage{changepage}

\DeclareMathAlphabet{\mathpzc}{OT1}{pzc}{m}{it}

\def\URL#1#2{\@ifundefined{href}{#2}{\href{#1}{#2}}}

\def\UrlOrds{\do\*\do\-\do\~\do\'\do\"\do\-}%
\g@addto@macro{\UrlBreaks}{\UrlOrds}

\edef\fntEncoding{\f@encoding}

\makeatother

\newif\ifmultipleabstract\multipleabstractfalse%
    \makeatletter
\def\ead{\@ifnextchar[{\@uad}{\@ead}}
\gdef\@ead#1{\bgroup
   \def\_{\string\underscorechar\space}
   \def\{{\string\lbracechar\space}
   \def\textdagger{\string\textdagger\space}
   \def\texttildeapprox{\string\texttildeapprox\space}
   \def~{\hashchar\space}
   \def\}{\string\rbracechar\space}
   \edef\tmp{\the\@eadauthor}
   \immediate\write\@auxout{\string\emailauthor
     {#1}{\expandafter\strip@prefix\meaning\tmp}}
  \egroup
}

\gdef\emailauthor#1#2{\stepcounter{ead}
      \g@addto@macro\@elseads{\raggedright
      \let\corref\@gobble
      \eadsep\texttt{#1} (#2)
      \def\eadsep{\unskip,\space}}
}

\makeatother
  
\let\citep\cite
\let\citet\cite    
  
\usepackage{float}

\newfloat{chart}{tbp}{Chart}
\floatname{chart}{Chart}
\makeatletter
\@ifpackageloaded{endfloat}{\@ifpackagewith{endfloat}{figuresonly}{\DeclareDelayedFloat{chart}{Chart}}{\@ifpackagewith{endfloat}{tablesonly}{}{\DeclareDelayedFloat{chart}{Chart}}}}{}\makeatother

\usepackage[numbers]{natbib}
\biboptions{numbers,compress}

\DeclareUnicodeCharacter{03C1}{$\rho$}

\RequirePackage{hyperref}%
\gdef\breakurldefns{%
\if@pdflatex\else%
  \RequirePackage[hyphenbreaks]{breakurl}%

\fi}%
\breakurldefns%

\hypersetup{%
        colorlinks,
        breaklinks=true,
        plainpages=false,%
        citecolor=blue,
        linkcolor=blue,
        urlcolor=blue,
        bookmarksopen=true,%
        bookmarksnumbered=false,%
        bookmarksdepth=5,%
        linktoc=page 
}

\usepackage{fancyhdr}
\fancypagestyle{footerwithlink}{
  \fancyhf{}

  \fancyfoot[R]{\footnotesize\thepage}

}

\fancypagestyle{plain}{%
  \fancyhf{}%

  \fancyfoot[R]{\footnotesize\thepage}%
}
\fancypagestyle{rotating}{%
  \fancyhf{}%

  \fancyfoot[R]{\footnotesize\thepage}%
}

\makeatletter
\newcommand{\FrontTitleFace}{%
\Large\fontseries{bx}\selectfont\raggedright
}
\newcommand{\NavTitleAuthors}{%
  {\FrontTitleFace \@title\par}%
  \vspace{.9\baselineskip}%
  {\normalfont\normalsize \elsauthors\par}%

}
\makeatother

\usepackage[figuresright]{rotating}

\makeatletter
\AtBeginDocument{\@ifpackageloaded{rotating}{\PassOptionsToPackage{figuresright}{rotating}}{\usepackage[figuresright]{rotating}}\setlength{\rotFPtop}{0pt plus 1fil}\setlength{\rotFPbot}{0pt plus 1fil}}
\makeatother

\makeatletter
 \AtBeginDocument{%
  \@ifpackagewith{endfloat}{figuresonly}
  {\DeclareDelayedFloatFlavor{sidewaysfigure}{figure}}
  {\@ifpackagewith{endfloat}{tablesonly}{\DeclareDelayedFloatFlavor{sidewaystable}{table}\DeclareDelayedFloatFlavor{longtable}{table}\DeclareDelayedFloatFlavor{landscape}{table}}
  {\@ifpackageloaded{endfloat}{\DeclareDelayedFloatFlavor{sidewaysfigure}{figure}\DeclareDelayedFloatFlavor{sidewaystable}{table}\DeclareDelayedFloatFlavor{longtable}{table}\DeclareDelayedFloatFlavor{landscape}{table}}{}}
  }
  }
\makeatother

\usepackage{float}      
\usepackage{rotfloat}   

\usepackage{needspace}  
\usepackage{changepage}
\usepackage{afterpage}

\usepackage[font=small,labelfont=bf]{caption}

\usepackage[font=small,labelfont=bf]{caption}

\usepackage{xparse} 

\NewDocumentCommand{\suppcaption}{O{} m}{%
  \begingroup
    \captionsetup{labelformat=empty,#1}%
    \caption{%
       \begin{minipage}[t]{\linewidth}
        \fontsize{11pt}{13.2pt}\selectfont
        #2
      \end{minipage}
    }%
  \endgroup
}

\NewDocumentCommand{\suppcaptionalm}{O{} m}{%
  \begingroup
    \captionsetup{labelformat=empty,#1}%
    \caption{%
      \begin{minipage}[t]{\linewidth}
        \fontsize{10pt}{12pt}\selectfont
        #2
      \end{minipage}
    }%
  \endgroup
}

\NewDocumentCommand{\suppcaptionmed}{O{} m}{%
  \begingroup
    \captionsetup{labelformat=empty,#1}%
    \caption{%
      \begin{minipage}[t]{\linewidth}
        \fontsize{9pt}{10pt}\selectfont
        #2
      \end{minipage}
    }%
  \endgroup
}

\NewDocumentCommand{\suppcaptionmid}{O{} m}{%
  \begingroup
    \captionsetup{labelformat=empty,#1}%
    \caption{%
      \begin{minipage}[t]{\linewidth}
        \fontsize{8pt}{9pt}\selectfont
        #2
      \end{minipage}
    }%
  \endgroup
}

\NewDocumentCommand{\suppcaptiontiny}{O{} m}{%
  \begingroup
    \captionsetup{labelformat=empty,#1}%
    \caption{%
      \begin{minipage}[t]{\linewidth}
        \fontsize{7pt}{8pt}\selectfont
        #2
      \end{minipage}
    }%
  \endgroup
}

\usepackage{url,multirow,morefloats,floatflt,cancel,tfrupee}

\usepackage{tcolorbox}
\definecolor{lightgray}{gray}{0.95} 
\definecolor{midgray}{gray}{0.7} 
\definecolor{darkgray}{gray}{0.1} 
\usepackage[svgnames]{xcolor}
\usepackage[x11names]{xcolor}
\usepackage{multicol}

\usepackage{etoolbox}
\makeatletter
\patchcmd{\@startsection}
  {\@afterindenttrue}
  {\@afterindentfalse}
  {}{}
\makeatother

\abstracttitle{Abstract}

\usepackage{xpatch}

\xpatchcmd{\MaketitleBox}{\hrule\vskip12pt}{\hrule\vskip12pt}{}{}

\xpatchcmd{\MaketitleBox}{\hrule\vskip40pt}{\hrule\vskip40pt}{}{}

\usepackage{helvet}

\makeatletter

\xpatchcmd{\MaketitleBox}{\itshape}{\normalfont}{}{}

\def\ps@pprintTitle{\save@ps@pprintTitle\gdef\@oddfoot{\footnotesize\normalfont \null\hfill\raisebox{-40pt}{\thepage}}}

\renewcommand{\ps@plain}{%
  \let\@oddhead\@empty
  \let\@evenhead\@empty
  \def\@oddfoot{\normalfont\hfil\thepage}%
  \let\@evenfoot\@oddfoot
}
\renewcommand{\ps@myheadings}{%
  \let\@oddhead\@empty
  \let\@evenhead\@empty
  \def\@oddfoot{\normalfont\hfil\thepage}%
  \let\@evenfoot\@oddfoot
}

\xpatchcmd{\MaketitleBox}
  {\Large\bfseries\raggedright}
  {\fontfamily{phv}\selectfont\LARGE\fontseries{bx}\selectfont\raggedright}
  {}{}

\makeatletter

\renewcommand\section{\@startsection{section}{1}{\z@}%
  {0.9\baselineskip}
  {0.5\baselineskip}
  {\normalfont\large\bfseries}}

\renewcommand\subsection{\@startsection{subsection}{2}{\z@}%
  {0.7\baselineskip}
  {0.3\baselineskip}
  {\normalfont\normalsize}}

\makeatother

\usepackage{ragged2e}

\newcommand{\metaheading}[1]{%
  \par
  \vspace{0.25\baselineskip}
  \noindent\textbf{#1}\par
  \vspace{0.1\baselineskip}
}

\usepackage{placeins}
\usepackage{xparse}
\usepackage{geometry}

\makeatletter
\newenvironment{shrinkPageWithFooter}[1][top=2.5cm,bottom=2.5cm,left=2.5cm,right=2.5cm]{%
  \begingroup%
  \newgeometry{#1}%
  \pagestyle{fancy}%
  \fancyhf{}%
  \fancyfoot[R]{\footnotesize\thepage}%
}{%
  \clearpage%
  \endgroup%
  \restoregeometry
  \clearpage%
}
\makeatother

\newcommand{\unnumberedsection}[1]{%
  \phantomsection
  \addcontentsline{toc}{section}{#1}%
  \section*{#1}%
}

\newcommand{\unnumberedsubsection}[1]{%
  \phantomsection
  \addcontentsline{toc}{subsection}{#1}%
  \subsection*{#1}%
}

\usepackage{placeins}

\makeatletter
\renewcommand\@makefntext[1]{%
  \parindent 0pt\relax
  \noindent
  \setbox0=\hbox{\@makefnmark\hspace{0.4em}}%
  \hangindent=\wd0 \hangafter=1\relax
  \box0 #1%
}
\makeatother

\usepackage{dblfloatfix}

\usepackage{cuted}

\begin{document}

\begin{frontmatter}

    \title{ \raggedright
  \textbf{Clinical utility of foundation models in musculoskeletal MRI for biomarker fidelity and predictive outcomes}
  }

\author[affd5e08f3c12cc4e4085348c07ab42dfa3,aff3f7f2a270cc54efa8955d6de83c489fc,affa1f10290191d4e52a7f7c8707c55d6b2]{\raggedright Gabrielle Hoyer\corref{contrib-edd8a42850c441099e6b68f87996d463}}
\ead{gabbie.hoyer@ucsf.edu}\cortext[contrib-edd8a42850c441099e6b68f87996d463]{Corresponding author.}
\author[affd5e08f3c12cc4e4085348c07ab42dfa3,aff3f7f2a270cc54efa8955d6de83c489fc,affa1f10290191d4e52a7f7c8707c55d6b2]{Michelle W~Tong}
\author[affd5e08f3c12cc4e4085348c07ab42dfa3,aff3f7f2a270cc54efa8955d6de83c48new]{Rupsa Bhattacharjee}
\author[affd5e08f3c12cc4e4085348c07ab42dfa3,aff8d61c628cea74863888a613b7e564205]{Valentina Pedoia}
\author[affd5e08f3c12cc4e4085348c07ab42dfa3,affa1f10290191d4e52a7f7c8707c55d6b2]{Sharmila Majumdar}

\address[affd5e08f3c12cc4e4085348c07ab42dfa3]{ \raggedright
    Center for Intelligent Imaging, Department of Radiology and Biomedical Imaging\unskip, University of California, San Francisco\unskip, CA\unskip, USA
    }
    
\address[aff3f7f2a270cc54efa8955d6de83c489fc]{
    Department of Bioengineering\unskip, 
    University of California, Berkeley\unskip, CA\unskip, USA
    }
  	
\address[affa1f10290191d4e52a7f7c8707c55d6b2]{
    Department of Bioengineering and Therapeutic Sciences\unskip, 
    University of California, San Francisco\unskip, CA\unskip, USA
    }

\address[aff3f7f2a270cc54efa8955d6de83c48new]{
    Department of Medical Sciences and Technology\unskip, 
    Indian Institute of Technology, Madras\unskip, Chennai\unskip, India
    }
    
\address[aff8d61c628cea74863888a613b7e564205]{
    Bay Area Institute of Computation\unskip, Altos Labs, Redwood City\unskip, CA\unskip, USA
    }

\clearpage


\begin{abstract}
\noindent
Precision medicine in musculoskeletal imaging requires scalable measurement infrastructure. We developed a modular system that converts routine MRI into standardized quantitative biomarkers suitable for clinical decision support. Promptable foundation segmenters (SAM, SAM2, MedSAM) were fine-tuned across heterogeneous musculoskeletal datasets and coupled to automated detection for fully automatic prompting. Fine-tuned segmentations yielded clinically reliable measurements with high concordance to expert annotations across cartilage, bone, and soft tissue biomarkers. Using the same measurements, we demonstrate two applications: (i) a three-stage knee triage cascade that reduces verification workload while maintaining sensitivity, and (ii) 48-month landmark models that forecast knee replacement and incident osteoarthritis with favorable calibration and net benefit across clinically relevant thresholds. Our model-agnostic, open-source architecture enables independent validation and development. This work validates a pathway from automated measurement to clinical decision: reliable biomarkers drive both workload optimization today and patient risk stratification tomorrow, and the developed framework shows how foundation models can be operationalized within precision medicine systems.

\end{abstract}
  \end{frontmatter}

\makeatletter
\let\ps@plain\ps@footerwithlink     
\let\ps@myheadings\ps@footerwithlink 
\let\ps@pprintTitle\ps@footerwithlink 
\makeatother
\pagestyle{footerwithlink}          

\setlength{\footskip}{36pt}

\setlength{\parindent}{0pt}
\setlength{\parskip}{0.6\baselineskip}


\section{Introduction}

AI in medical imaging is shifting from narrow benchmarks to systems that convert routine scans into standardized, quantitative measurements linked to clinical decisions. In radiology, triage tools already show that time to interpretation can be reduced when algorithms are embedded directly into workflow \cite{annarumma_automated_2019, oneill_active_2021, batra_radiologist_2023}. Concurrently, quantitative MRI (qMRI) and radiomics provide reproducible biomarkers that enable predictive modeling for personalized care \cite{gillies_radiomics_2016, eckstein_quantitative_2011}.

Musculoskeletal disorders are a leading source of disability worldwide and are projected to rise with population aging. These conditions disproportionately affect low- and middle-income settings, which makes musculoskeletal (MSK) imaging a particularly important domain for precision medicine solutions that can improve both healthcare access and outcomes \cite{cieza_global_2020, hartvigsen_what_2018, williams_musculoskeletal_2018}.

Despite this need, routine MSK radiology workflows rarely produce quantitative biomarkers from MRI. Doing so often requires trained experts to delineate multiple structures across hundreds of slices (e.g., cartilage, bone, muscle, intervertebral discs), followed by quality control. This time- and expertise-intensive process, combined with inter-reader variability in low-contrast or degenerated anatomy, limits scalability in clinical practice and motivates automated pipelines that can generate measurements reliably across diverse acquisition conditions.

Standardized measurement is further complicated by heterogeneity in MRI protocols across scanners, sites, joints, and tissues \cite{tajbakhsh_embracing_2020}.


\begin{shrinkPageWithFooter}[top=2.5cm,bottom=2.5cm]

\begingroup
  \fixFloatSize{images_final/main_fig_1_small.pdf}
  \begin{figure*}[!htbp]

\captionsetup{position=above, singlelinecheck=false, justification=raggedright}
\centering
\caption[]{\textbf{Musculoskeletal MRI segmentation study design.}}
\makeatletter

\IfFileExists{images_final/main_fig_1_small.pdf}
  {\includegraphics[width=1.0\linewidth]{images_final/main_fig_1_small.pdf}}
  {\includegraphics[width=1.0\linewidth]{main_fig_1_small.pdf}}

\makeatother

\captionsetup{position=below}

\begin{center}
\begin{minipage}{\linewidth}
\small\justify
\hyphenation{anatomy musculo-skeletal segmen-tation bio-markers evalua-tion}
\textbf{a)} Overview of the musculoskeletal anatomies represented in the study datasets, including shoulder, spine, hip, thigh, and knee. These anatomies reflect the clinical use cases covered by the segmentation models. The panel also introduces the experimental framework for foundation model development, including baseline evaluation, dataset ablation, strategic dataset mixing, and fine-tuning approaches,culminating in the creation of a musculoskeletal (MSK) foundation model. \textbf{b)} Design for clinical biomarker computation and evaluation. Segmentation outputs are used to derive imaging biomarkers (e.g., cartilage thickness, muscle volume, disc height, T\ensuremath{_{1\ensuremath{\rho}}}/T\ensuremath{_{2}} relaxation times), which are then compared against ground truth annotations using statistical methods including regression analysis, Bland-Altman comparison, and intraclass correlation coefficients. \textbf{c)} Scalable, fully automatic segmentation pipeline. A detection model generates slice-wise bounding box prompts that pass to a SAM-variant model. \textbf{d)} Batch segmentation and biomarker extraction feeds the triage case routing and survival/ risk model pipelines introduced in \hyperref[fig:clinical]{Fig.~\ref*{fig:clinical}} and \hyperref[fig:landmark]{Fig.~\ref*{fig:landmark}}.
\end{minipage}
\end{center}
\label{fig:overview}
\end{figure*}
\endgroup

\end{shrinkPageWithFooter}
\FloatBarrier
\clearpage


 In this context, segmentation serves as the measurement interface for qMRI and produces biomarkers such as cartilage thickness maps, T\ensuremath{_{1\ensuremath{\rho}}}/T\ensuremath{_{2}} relaxation times, disc height, and muscle volume that support both population-scale research and patient-specific predictions \cite{eckstein_quantitative_2011, pons_quantifying_2018, tunset_method_2013}. To address these barriers, we employ foundation segmentation models (SAM, MedSAM, SAM2) combined with automated bounding-box prompting via object detection. The system reduces annotation burden while preserving measurement accuracy across diverse anatomies \cite{Kirillov_2023_ICCV_SegmentAnything, ravi2025sam, ma_segment_2024_medsam}. This modular architecture insulates clinical decision support from segmentation model turnover. Sustainable healthcare deployment depends on stability at the point of clinical action rather than at the model level \cite{radiology_technical_committee_integrating_2025, leiner_bringing_2021, brink_acrs_2022}.

Here we present a musculoskeletal MRI framework spanning five anatomies (knee, hip, shoulder, lumbar spine, thigh) and validated across 12 datasets totaling 913 scans from heterogeneous clinical and research MRI protocols. Across all five anatomies, we evaluate segmentation and biomarker agreement. The system establishes how foundation models, once fine-tuned, can generate clinically validated biomarkers that enable distinct precision medicine applications; we illustrate downstream translation on knee MRI, where large labeled cohorts enable end-to-end evaluation at scale (\hyperref[fig:overview]{Fig.~\ref*{fig:overview}}). The first application is an automated triage cascade for knee MRI. The cascade uses cartilage and bone measurements to flag studies for review, potentially reducing workload without sacrificing sensitivity for pathology detection. Second, a predictive modeling framework uses longitudinal cartilage and meniscus thickness trajectories from serial MRIs in the Osteoarthritis Initiative to forecast osteoarthritis (OA) progression and estimate individual risk of knee replacement at clinically relevant time horizons \cite{eckstein_quantitative_2013, kwoh_predicting_2020, van_houwelingen_dynamic_2011, van_houwelingen_dynamic_2008}. 

With segmentation as the measurement interface, we validate biomarker fidelity against expert measurements and demonstrate utility through two clinical applications: (i) operational triage with explicit workload targets and (ii) longitudinal risk modeling via landmark methods, with utility quantified by decision-curve analysis across clinically relevant thresholds \cite{van_houwelingen_dynamic_2011, van_houwelingen_dynamic_2008, vickers_decision_2006}. This modular approach supports both predictive modeling and vendor-neutral deployment \cite{radiology_technical_committee_integrating_2025, leiner_bringing_2021, brink_acrs_2022}, while positioning routine MSK MRI as quantitative input that can integrate with other data streams (e.g., gait, wearables, health records) for proactive care solutions. The study follows this automation-to-measurement-to-decision framework: first, establishing biomarker reliability, then demonstrating how validated measurements enable both immediate operational benefits and longitudinal risk prediction.


\section{Methods}

\subsection{MRI Datasets and Annotations}

We assembled 12 musculoskeletal MRI datasets spanning knee, hip, lumbar spine, shoulder, and thigh anatomy, totaling 913 scans with expert tissue annotations (Supplementary Table S1). The cohort includes five knee datasets (497 scans) \cite{tolpadi_k2s_2023, pedoia_principal_2016, peterfy_osteoarthritis_2008, noauthor_white_2017}, four lumbar spine datasets (296 scans) \cite{hess_deep_2023}, and one dataset each for thigh (50 scans) \cite{peterfy_osteoarthritis_2008}, hip (42 scans) \cite{thahakoya_rafeek_evaluating_2023}, and shoulder (28 scans) \cite{lee_magnetic_2015, nardo_quantitative_2014}. Acquisitions span 3D research and 2D clinical protocols (spin-echo, gradient-echo, CUBE, DESS, quantitative T\textsubscript{1ρ}/T\textsubscript{2}) at 1.5T and 3T on Siemens or GE systems.

Except for the multi-center Osteoarthritis Initiative (OAI) cohort, datasets were acquired within the UCSF health system across multiple imaging facilities and scanner platforms. Thus, while sometimes described as single-site, the UCSF collection reflects multi-facility and multi-vendor variability within one health system (Supplementary Table S2). Dataset intent and protocol mix varied across anatomy: the lumbar spine and thigh cohorts are routine 2D clinical spin-echo acquisitions, whereas several knee, hip, and shoulder cohorts use high-resolution 3D and quantitative protocols originally collected for research aims.

Seven of the 12 datasets include 50 scans or fewer; accordingly, we emphasize uncertainty intervals and tail summaries and interpret per-dataset hypothesis tests descriptively. The cohorts were not assembled to stress-test severe motion, substantial susceptibility from hardware, or post-operative implants. Performance under these conditions should be evaluated before clinical use.

Key acquisition parameters (TE, TR, flip angle, voxel spacing) and label definitions appear in Supplementary Tables \hyperref[fig:supp-s0]{S0}--S2 (Supplementary Information 5.1). These datasets support tasks 


\bgroup
\fixFloatSize{images_final/main_fig_2.png}
\begin{figure*}[!htbp]

\captionsetup{position=above, singlelinecheck=false, justification=raggedright}
\centering
\caption[]{\textbf{Summary of dataset composition, subject demographics, and imaging protocol characteristics.}}
\makeatletter
\IfFileExists{images_final/main_fig_2.png}
  {\includegraphics{images_final/main_fig_2.png}}
  {\includegraphics{main_fig_2.png}}
\makeatother

\captionsetup{position=below}

\begin{center}
\begin{minipage}{\linewidth}
\small\justify
\hyphenation{anatomy groups musculo-skeletal segmen-tation distri-bution charac-teristics}
\textbf{a)} Age distribution across the five musculoskeletal anatomy groups (shoulder, hip, knee, lumbar spine, and thigh), shown as kernel density plots. \textbf{b)} Weight distribution for the same anatomy groups, plotted using the same format. \textbf{c)} Sex distribution (male vs. female) within each anatomy group, displayed as donut charts with corresponding percentages and subject counts. \textbf{d)} Distribution of segmentation labels across the dataset. The inner ring displays the proportion of tissue types (cartilage, bone, muscle, nerve, fat), while the outer ring shows their anatomical origin. \textbf{e)} Upset plot illustrating the overlap of key MRI acquisition parameters, including scanner field strength, slice thickness, TR, TE, flip angle, 3D acquisition, and vendor. \textbf{f)} Number of segmented slices available for each anatomy group (log scale). \textbf{g)} Total number of unique subjects included per anatomy group.
\end{minipage}
\end{center}
\label{fig:demographics}
\end{figure*}
\egroup

\FloatBarrier


such as cartilage degeneration, muscle composition, reconstruction testing, and risk modeling.

Reference annotations were derived from prior studies that defined labeling protocols and performed quality checks (cited per dataset in Supplementary Information 5.1). Depending on the dataset, labels include bone, cartilage and menisci, muscle and fat compartments, intervertebral discs, and nerves (including the spinal nerve bundle) (Supplementary \hyperref[fig:supp-s0]{Table S0}). Repeat reads and multi-reader segmentations were not available for every dataset in the present study, so we did not compute new inter- or intra-reader variability analyses in this manuscript. Instead, we report biomarker agreement between model-derived and expert-derived measurements as an end-to-end validation signal.

Each dataset was originally collected for a specific clinical or research aim, resulting in heterogeneous contrasts and label taxonomies that support evaluation across multi-label and multi-instance tasks.


\subsection{Ethics and approvals}
This study was conducted in accordance with the Declaration of Helsinki (1964) and its subsequent amendments, as well as all relevant regulations. Data acquisition and machine learning analyses were approved by the University of California, San Francisco (UCSF) Institutional Review Board (IRB), operating under Federalwide Assurance \#00000068, with specific IRB approvals including 21-33865, 19-29744, 17-22581, 13-11605, and 13-11671 (data acquisition), and 18-24775 (machine learning use), respectively. The Osteoarthritis Initiative (OAI) study was conducted in accordance with IRB approvals at each OAI clinical site, including Memorial Hospital of Rhode Island, Ohio State University, University of Pittsburgh, and University of Maryland/Johns Hopkins University, with the OAI Coordinating Center at UCSF providing IRB approval (approval number 10-00532, Federalwide Assurance \#00000068) and the OAI Clinical Sites Single IRB of Record approved as study number 2017H0487, Federalwide Assurance \#00006378. This study is registered with ClinicalTrials.gov (identifier: NCT00080171), and an independent Observational Study Monitoring Board (OSMB) appointed by the National Institute of Arthritis and Musculoskeletal and Skin Diseases (NIAMS) oversaw adherence to ethical research standards and participant safety. All studies, including those involving UCSF data acquisition, machine learning analyses, and the OAI study, were funded by NIAMS, and all participants provided written informed consent prior to participation.


\subsection{Model Selection and Configuration}
We evaluated three foundation segmentation models to assess prompt-based MSK MRI segmentation within a scalable framework: SAM ViT-B \cite{Kirillov_2023_ICCV_SegmentAnything}, MedSAM ViT-B \cite{ma_segment_2024_medsam}, and SAM2 \texttt{sam2\_hiera\_base\_plus} \cite{ravi2025sam}. We selected base-sized variants with broadly comparable parameter counts to reduce confounding by model capacity and to keep inference and training costs within realistic clinical research infrastructure constraints (e.g., workstation- or server-class GPUs). Controlling for model scale also reflects practical constraints for prospective deployment, where latency and hardware availability can be limiting.

SAM (ViT-B) ($\sim$91M parameters) was trained on $\sim$1B natural image-mask pairs and uses an image encoder, prompt encoder, and mask decoder for promptable segmentation. MedSAM (ViT-B) adapts SAM to medical imaging, trained on $\sim$1.5M medical image-mask pairs across modalities and anatomies while retaining the same promptable design. SAM2 (\texttt{sam2\_hiera\_base\_plus}, $\sim$81M parameters) employs a hierarchical multi-scale encoder trained on the SA-V video dataset (50{,}900 videos; 642{,}600 masklets; $>$35M masks). We evaluated SAM2 with 2D prompts only to remain consistent with SAM/MedSAM and to avoid long-sequence memory issues in slice-rich MRI volumes.

We limited model comparisons to promptable foundation segmenters with public weights and a common box-prompt interface, because the study goal was to evaluate an end-to-end measurement pipeline that can swap segmentation backbones without changing biomarker extraction. Conventional fully supervised 3D segmentation networks (e.g., 3D U-Net variants) were not included as baselines here, because benchmarking them across all anatomies would require matched supervision, architecture selection, and tuning across many tasks beyond the scope of this work.


\subsection{Prompting strategy (2D prompts)}
Slice-wise bounding-box prompts were generated from manual masks and applied in every experiment, including AutoLabel tests. We standardized on uniform 2D boxes across anatomy, sequence, and scanner to enable consistent prompting and high-throughput inference. For SAM2, we disabled the memory module because long, slice-rich MRI volumes induce label occlusion/entry/exit and repeated re-prompting, which undermines automation and consistency. Evaluating all models with independent 2D prompts isolates anatomy and training effects from model-specific tracking behaviors, and ensures consistent scaling across diverse patient cohorts. Additional pilot observations are provided in Supplementary Information 5.2 (``Prompts and automation rationale;'' see 5.2.2 for the rationale for disabling SAM2 memory).


\subsection{Image Preprocessing}
A YAML-driven Python pipeline standardized all scans (intensity normalization, resampling to 1{,}024\(\times\)1{,}024, RGB stacking) and stored transform maps for inverse projection. A 256\(\times\)256 path, matching SAM/SAM2 logit size, supported data-efficiency trials. Masks were resized with nearest-neighbor interpolation and cleaned with one morphological closing pass. Full workflow and metadata handling are provided in Supplementary Information 5.3 (\enquote{Preprocessing Strategy for Model Compatibility;} 5.3.1–5.3.5) and 6.1.2 (\enquote{Metadata Integration and Imaging Fidelity}).


\subsection{Dataset Partitioning for Balanced Representation}
Datasets were partitioned into training (70\%), validation (15\%), and test (15\%) subsets. Subject-level splits were stratified by sex to help ensure balanced demographic representation across experimental conditions. Sex was chosen because it is consistently available across datasets and is associated with differences in musculoskeletal morphology and body composition; stratification reduces the chance that demographic imbalance drives apparent performance differences.

Additional variables such as age, BMI, or clinical outcomes can be used for stratification when required by study-specific aims.


\subsection{Segmentation metrics and statistical framework}
We quantified segmentation with the Dice similarity coefficient \cite{dice_measures_1945} and the Jaccard index (intersection-over-union) \cite{jaccard_etude_1901}, computed per slice, then aggregated to subject and dataset levels via a unified evaluation pipeline. Both metrics range from 0 (no overlap) to 1 (perfect overlap). Structure-level metrics were aggregated using the median to ensure robust summary statistics, except where noted.

We additionally report lower-tail subject-level Dice for each dataset and model configuration (median, 5th percentile, and minimum across subjects) to complement central-tendency summaries and better reflect worst-case behavior. These tail summaries are provided in Supplementary Data D55 (all experiments) and in compact forms aligned to the models shown in \hyperref[table:tiered_table]{Table~\ref*{table:tiered_table}} (Supplementary Data D56--D57). Because several datasets have modest test-set sizes, these tail statistics are intended as descriptive indicators of reliability rather than as stand-alone inferential endpoints.

For model comparisons across the same subjects, we used a Friedman test (non-parametric repeated measures)~\cite{friedman_use_1937} followed by Wilcoxon signed-rank pairwise tests~\cite{wilcoxon_individual_1945} with Benjamini-Hochberg false discovery rate (FDR) control~\cite{benjamini_controlling_1995} (5\%); all tests were two-sided with $\alpha=0.05$. We report test statistics, raw $p$ values, and FDR-adjusted $p$ values.

Extended statistics and per-dataset tables are provided in Supplementary Tables S3--S5, and class-level distributions in Supplementary Figs. S1--S4; the evaluation pipeline is detailed in Supplementary Information 6.2.


\subsection{Baseline (zero-shot) assessment}
We evaluated three promptable Vision Transformer segmenters (SAM ViT-B, SAM2 sam2\_hiera\_base\_plus, and MedSAM ViT-B fine-tuned for medical images) under a consistent bounding-box prompt regime derived from manual masks. The prompting configuration matched that used for fine-tuning experiments (see \enquote{Prompting strategy}). Comparative testing followed the framework above (Friedman, then Wilcoxon with BH FDR 5\%). Summary statistics reside in Supplementary Tables S3–S5; extended baseline detail appears in Supplementary Information 5.4.


\subsection{Fine-tuning configuration and training setup}

We assessed clinical readiness across heterogeneous MSK MRI using four protocolized scenarios. These included: (i) data-efficiency tiers of 5, 10, 20, 40, and full subjects, chosen to span a practical range of small-to-moderate annotation budgets in approximately doubling steps; (ii) architecture ablation via encoder frozen versus unfrozen with bounding-box shift augmentation ($\pm$20 px); (iii) dataset mixing by anatomy/sequence (knee; spine; muscle-focused thigh$+$spine); and (iv) a pooled mskSAM run that trained on every dataset except a held-out knee cohort (experiment index in Supplementary \hyperref[fig:supp-s6]{Table~S6}). Extended ablation details are provided in Supplementary Information 5.5.

SAM participated in all scenarios. SAM2 and MedSAM were fine-tuned only in the pooled run, reflecting compute constraints and their architectural similarity to SAM.

A YAML-driven pipeline applied consistent preprocessing for intensity, spacing, and shape (resampling to $1024\times1024$ with RGB stacking) and preserved transforms for inverse projection; a $256\times256$ logit path matched SAM/SAM2 internals for data-efficiency trials. Masks used nearest-neighbor resizing with one morphological closing pass.

Unless stated otherwise, we used AdamW (learning rate $1\times10^{-4}$, weight decay $0.01$)~\cite{loshchilov_decoupled_2017_2019} with a CosineAnnealingWarmRestarts schedule~\cite{loshchilov_sgdr_2016_2017}; early stopping (patience 10, $\Delta=10^{-4}$); gradient clipping (max-norm 1.0) and accumulation (4 steps); mixed-precision (CUDA Amp)~\cite{nickolls_scalable_2008}; batch size 2; and a compound loss (Dice $+$ BCEWithLogits).

We compared full fine-tuning (encoder$+$decoder) with partial fine-tuning (decoder only) to balance domain adaptation against compute cost. Bounding-box shift augmentation (uniform jitter up to $\pm$20 px in the resized $1024\times1024$ model input space; $\approx$2\% of image width/height) replicated the prompt variability inherent in both automated detection and manual annotation workflows. Because inputs are resized to a fixed resolution before inference, deployments using different preprocessing resolutions can scale the jitter proportionally (e.g., $\pm0.02\times$image width/height).

Single-dataset and mixed configurations were trained for knees (MAPSS echo1, 3D DESS, 3D CUBE), spine (T$_1$ axial/sagittal, T$_2$ axial), and a muscle-focused mix (T$_1$ axial thigh$+$spine). The pooled mskSAM included all datasets except the 8$\times$ undersampled knee CUBE (held out for robustness testing).

To study data efficiency, we trained on 5/10/20/40-subject tiers (subject-based rather than slice-based) across single- and multi-dataset settings; where fewer subjects existed, we used the maximum available subset. We used logarithmically spaced tiers to capture rapid early gains in low-data regimes while limiting compute overhead, which is standard in data-efficiency evaluations.

We also compared initial weights from SAM, MedSAM, and SAM2 under matched configurations across knee, spine, shoulder, and thigh in both full/partial settings (summary in Supplementary Table S9; class-level metrics in Supplementary Data D1--D24).

Slice- and subject-level numerical results are provided in Supplementary Data D1--D26; tissue- and label-level breakdowns are in Supplementary Tables S7--S9 and Supplementary Figs. S1--S4.


\subsection{Hierarchical Mixed-Effects Modeling}

We used hierarchical linear mixed-effects models to quantify how MRI acquisition parameters and experimental factors relate to segmentation quality and to inform deployment. The dependent variable was the subject-level mean Dice.

Because the same subjects contribute predictions under multiple experimental conditions and are nested within datasets acquired under distinct protocols, simple regression would violate independence assumptions. A mixed-effects specification addresses within-subject correlation and between-dataset heterogeneity. This analysis estimates overall trends that inform system configuration, rather than establishing causal effects of specific imaging parameters. Full preprocessing steps and model specification appear in Supplementary Information 5.6.

We assembled a modeling matrix from MRI acquisition metadata (e.g., echo time [TE], repetition time [TR], flip angle, field strength, scanner vendor, pixel spacing, slice thickness, image row size, SAR) and experimental covariates (acquisition mode, encoder fine-tuning, prompt shift). Continuous variables were standardized; categorical nominal variables were one-hot encoded with a dropped reference level; ordinal variables were ordinally encoded. Missing values were imputed by iterative random-forest models (regressor/classifier)~\cite{stekhoven_missforestnon-parametric_2012}. Imputation fidelity was checked by comparing observed and imputed distributions using Kolmogorov-Smirnov tests (Supplementary Table S10).

To reduce multicollinearity, variance inflation factors (VIF)~\cite{kutner_applied_2005} were computed on the preprocessed matrix; features with VIF > 10 were removed and VIFs recomputed to confirm reduction. The final retained feature set for each analysis is listed in Supplementary Table S11.

Separate models were fitted for each fine-tuning regime (Type 1: single-dataset; Type 2: grouped by anatomy/sequence; Type 3: pooled mskSAM). Fixed effects comprised TE, pixel spacing, slice thickness, flip angle, acquisition mode (2D$=$0, 3D$=$1), image-encoder fine-tuned (0/1), and bounding-box shift applied (0/1). Dataset identity entered as a random intercept to account for between-dataset heterogeneity. Where appropriate, two-way interactions were included to test whether imaging-effect relationships differed by regime. Models were estimated by restricted maximum likelihood; 95\% confidence intervals and p-values used Wald approximations. Complete coefficient tables, interaction terms, diagnostics, and per-scenario outputs are provided in Supplementary Table S12 and Supplementary Data D25–D26.

Given heterogeneity in anatomy, sequences, tissue labels, and task difficulty across datasets, residual confounding of parameter effects with task mix is possible; the dataset random intercept mitigates but cannot eliminate this. Accordingly, p-values are treated as descriptive, and we emphasize effect sizes and 95\% CIs. The analysis informs protocol guardrails and model configuration choices rather than prescriptive scanner-parameter changes.

\hyperref[fig:biomarkers]{Fig.~\ref*{fig:biomarkers}a} summarizes fixed-effect estimates across regimes; Full numerical results and diagnostics are in the Supplementary materials cited above.


\subsection{Imaging Biomarkers and Agreement Analysis}
We derived quantitative measures from manual and model masks for four established biomarkers (cartilage thickness, intervertebral disc height, muscle volume, and mean T$_{1\rho}$/T$_2$ relaxation time) and two proxies (meniscus thickness and bone volume used to extend coverage and stress-test automation). Meniscus thickness used the same medial-axis distance-based pipeline as cartilage thickness.

For menisci, thickness corresponds to the same medial-axis distance-based estimate applied to the meniscus mask. For bone, we report bone volume computed as the segmented voxel count multiplied by voxel volume (reported in cm$^3$). Here, ``biomarker'' denotes any quantitative MRI measurement computed from a segmentation mask; proxies are not positioned as disease endpoints. All measurements were converted to millimeters, cubic centimeters, or milliseconds using DICOM (Digital Imaging and Communications in Medicine) metadata (PixelSpacing and SliceThickness); subject-level summaries for each anatomy-dataset pair are provided in Supplementary Data D27--D52.

Cartilage thickness was computed by applying a medial-axis transform to each mask, then measuring the Euclidean distance to the nearest boundary. Thickness values were sampled along the medial-axis skeleton, defined as the centerline equidistant from opposing cartilage boundaries, and averaged within each anatomical compartment, then scaled to physical units (Supplementary Fig.~S6).

Intervertebral disc height used connected components to identify disc instances. For each slice we computed the minimal bounding rectangle of the disc contour and took the cranio--caudal extent as slice-level height; the subject-level disc height per level was the maximum across slices. This definition is consistent with prior segmentation-based disc-height measurement approaches \cite{hess_deep_2023} (details and additional citations in Supplementary Information 5.7). Unlike cartilage thickness, which is a local laminar thickness estimated along a medial axis, disc height is intended as a global cranio--caudal extent, making a rectangle-based estimate a simple and robust choice.

Muscle and bone volumes were obtained by multiplying slice-wise voxel counts within each tissue mask by the voxel volume
$v=\mathrm{PixelSpacing}_x \times \mathrm{PixelSpacing}_y \times \mathrm{SliceThickness}\;(\mathrm{mm}^3)$,
summed across slices, and reported in cm$^3$.

For T$_{1\rho}$/T$_2$ mapping, cartilage masks were applied to parametric maps; voxel values were clipped to 0--100\,ms and averaged per compartment, then aggregated to the subject level.

Agreement between model-derived and manual biomarker values was assessed at the subject level. Distributional assumptions were screened with Shapiro-Wilk~\cite{shapiro_analysis_1965} (normality) and Levene~\cite{brown_forsythe_1974_levene} (homogeneity). When assumptions held, we computed an absolute-agreement, two-way mixed-effects single-measurement intraclass correlation coefficient (ICC(3,1))~\cite{koo_guideline_2016} with 95\% confidence intervals. Otherwise, we estimated a mixed-effects ICC from a linear mixed model with subject as a random effect and obtained 95\% intervals by bootstrap. Bland-Altman~\cite{bland_statistical_1986} analyses (parametric or percentile-based non-parametric, matched to assumptions), regression (ordinary least squares with $R^2$ when parametric; otherwise Gaussian process regression~\cite{rasmussen_gaussian_2005} with radial basis function (RBF) + White kernels and 95\% pointwise predictive intervals), and Spearman's $\rho$ complemented ICC to characterize bias and concordance. All tests were two-tailed with $\alpha=0.05$; where multiple comparisons were made across biomarker-dataset combinations, $p$ values were adjusted with Benjamini-Hochberg FDR (5\%). Confidence intervals and distribution checks used 10{,}000 bootstrap resamples at the subject level.

Extended extraction specifics, assumption checks, and supplementary analyses are in Supplementary Information 5.7 (Supplementary Tables \hyperref[fig:supp-s13]{S13}--S16; Supplementary Figs. S5--S10).


\subsection{AutoLabel Workflow Evaluation}

AutoLabel couples a YOLOv8m ~\cite{jocher_ultralytics_2023} detector with SAM-family segmenters (SAM, SAM2, MedSAM) to enable population-scale deployment without manual annotation. Detector-derived bounding boxes replace manual prompts for large-scale tests. We evaluated five segmentation configurations on five representative musculoskeletal datasets: \textit{Thigh\_2D\_T1ax\_Clinical\_Anatomical\_50}, \textit{Shoulder\_3D\_CUBE\_Research\_Anatomical\_28}, \textit{Spine\_2D\_T1sag\_Clinical\_Anatomical\_88}, \textit{Knee\_3D\_DESS\_Research\_Anatomical\_86}, and \textit{Spine\_2D\_T1ax\_Clinical\_Anatomical\_59}. These sets were chosen to span anatomy, contrast, and 2D/3D acquisition. The five model settings included zero-shot SAM, MedSAM, and SAM2, plus the best-performing fine-tuned model for each dataset, with and without bounding-box shift augmentation (details in Supplementary Information 5.8 and 6.2).

Prediction refinement used a standardized post-processing pipeline prior to evaluation: logits $\rightarrow$ sigmoid with a 0.5 threshold, small-component removal by connected components, a single morphological closing pass, and light boundary smoothing. Kernel sizing and any class-specific deviations are documented in Supplementary Information 5.8.1. Subject-level Dice was computed from slice-wise Dice and then aggregated to the dataset level; full per-dataset tables appear in Supplementary Data D53–D54.

We computed subject-level Dice under two conditions (ground-truth boxes and YOLO-derived boxes) and applied two-sided pairwise Wilcoxon rank-sum tests with Benjamini-Hochberg false-discovery control at 5\% to compare prompt sources (comparisons specified per dataset and model pairing). The unit of analysis ($n$) was the number of subjects per dataset. Visualization used raincloud plots to summarize distributions; complete statistics and figures are reported in Supplementary Table~S17 and Supplementary Figs.~S11–S15.


\subsection{Automated Knee MRI Multi-Stage Triage}\label{sec:triage}

We applied the AutoLabel $\rightarrow$ biomarker $\rightarrow$ classifier pipeline to 930 proton-density 3D CUBE knee scans (0.6\,mm isotropic voxels, eight-channel coil). AutoLabel (mskSAM2) produced femoral, tibial, and patellar cartilage masks and subchondral bone masks. Mean cartilage thickness and bone volume were z-scored to a fixed healthy reference defined as knees labeled normal for both bone and cartilage (bone\_label\,{=}\,0; cart\_label\,{=}\,0). For each biomarker, we subtracted the reference mean and divided by its sample standard deviation (SD); the reference was computed once globally (not per fold). Sex was encoded 0/1; cases with missing sex (<1\%) were excluded. Age and weight were used as recorded.

The feature set comprised six z-scored biomarkers (cartilage thickness and bone volume for femur, tibia, and patella) plus sex, age (years), and weight (kg).
 
 \textit{Stage A: normal-knee screen.} Elastic-net logistic regression ($\ell_1$ ratio 0.5; class-balanced)~\cite{zou_regularization_2005}, XGBoost~\cite{chen_xgboost_2016}, and histogram gradient boosting were trained with five-fold \emph{subject-grouped} cross-validation (CV). Out-of-fold probabilities were stacked with a logistic meta-learner. A 90\% specificity threshold (probability $\geq 0.938$) defined the Stage-A operating point.
 
 \textit{Stage B: cartilage \ensuremath{+} bone filter.} Using the Stage-A pass set, the same base learners and stacker were trained (five-fold grouped CV). Prespecified forwarding thresholds targeted 85\% (probability $\geq 0.893$) and 90\% (probability $\geq 0.911$) specificity.
 
 \textit{Stage C: joint and tissue localization.} We averaged logistic regression and XGBoost (equal weight). Part C1 labeled joint-level abnormality (femur, tibia, patella). Part C2 labeled tissue-level abnormality (bone vs cartilage) within each joint. Calibration curves for Stage C used 10 equal-width bins via \texttt{calibration\_curve}~\cite{pedregosa_scikit-learn_2011}; no probability recalibration was applied.

Five-fold grouped cross-validation (group\,{=}\,subject) produced out-of-fold predictions for Stages A and B; thresholds were set a priori at the specificity targets above. Stage C models were fit on the routed subset and evaluated without additional cross-validation.

We report area under the receiver operating characteristic (ROC) curve (AUC) and sensitivity at the preset specificity targets. Ninety-five percent confidence intervals (CIs) were computed via a 2,000-draw non-parametric bootstrap (resampling subjects with replacement). Confusion counts used \texttt{sklearn.metrics.confusion\_matrix} at the Stage-B thresholds; negative predictive value  NPV\,{=}\,TN/(TN\ensuremath{+}FN), and positive predictive value PPV\,{=}\,TP/(TP\ensuremath{+}FP), where TN, FN, TP, and FP denote true negatives, false negatives, true positives, and false positives. Stage-wise counts and calibration plots are provided in Supplementary Tables \hyperref[fig:supp-s18]{S18}--S19 and Supplementary Fig. S16; extended triage methods and runtime measurement appear in Supplementary Information 5.9 (5.9.1).

Verification workload was estimated as forwarded\_fraction $\times$ per-exam minutes and reported per 1{,}000 scans; for example, $99/930 \times 1{,}000 \times 2$ min $=$ 212.9 min $=$ 3.5 h at the 85\% setting, and $47/930 \times 1{,}000 \times 2$ min $=$ 101.1 min $=$ 1.7 h at the 90\% setting.

Throughput was characterized by timing AutoLabel on 30 randomly sampled knee 3D CUBE scans after a two-scan warm-up. \enquote{Model compute} comprised detector, SAM2 segmentation, and per-slice biomarker updates; file I/O and end-run aggregation were timed separately and excluded from model-compute figures. Raw per-scan timings (CSV) and per-component breakdowns are provided as Supplementary Data.


\subsection{Longitudinal Biomarker Extraction and Outcome Modeling}
We processed five visits per knee (baseline, 12, 24, 36, and 48 months) in 1{,}109 Osteoarthritis Initiative~\cite{peterfy_osteoarthritis_2008} index knees after MRI quality checks; surgery was tracked to 96--120 months. Cohort identification and visit selection followed our previously published protocol~\cite{hoyer_foundations_2025}. AutoLabel (mskSAM2) generated cartilage and meniscus masks at each visit; mean thickness and annual change were computed per compartment and z-standardized using the mean and standard deviation from knees with baseline Kellgren--Lawrence grades 0--1 (radiographically normal/possible early OA) as a reference distribution. This standardization does not transform KL grades; it only rescales imaging-derived biomarkers for comparability across compartments and time.

Here, the KL 0--1 group serves only as a fixed normalization reference for z-scoring; outcome modeling and evaluation are performed on the full at-risk landmark cohorts defined by the study endpoints (Supplementary Information 5.10.1).

Age, sex, and body mass index (BMI) were included; missing demographics were forward-filled when present ($<$3\%). Baseline demographics and landmark event rates for the OAI longitudinal analyses are summarized in Supplementary \hyperref[fig:supp-s20]{Table~S20} (TKR task) and Supplementary Table S22 (incident radiographic OA task).

A 48-month landmark~\cite{van_houwelingen_dynamic_2011} defined time $t_0$. At $t_0$, the risk set included knees uncensored and event-free with complete biomarkers through 48 months; models predicted outcomes over the subsequent 48 months. Two prediction tasks were studied: total knee replacement (TKR) (48$\rightarrow$96 months as the main horizon) and incident radiographic osteoarthritis (OA) (48 months post--$t_0$ in knees with baseline KL $<$ 2).

For TKR we used a 400-tree random forest~\cite{breiman_random_2001} (\texttt{min\_samples\_leaf}=10) with a logistic-regression baseline; for OA we used L2-regularized logistic regression. All models used biomarker trajectories available up to $t_0$ (thickness levels and annual change) plus age, sex, and BMI, with class weights set to ``balanced.''

Five-fold \texttt{StratifiedGroupKFold} (group=participant) preserved paired knees. Out-of-fold probabilities were calibrated with isotonic regression~\cite{zadrozny2002transforming}; discrimination and decision curves used the calibrated probabilities. Area under the ROC curve (AUC) was summarized with percentile bootstrap (1{,}000 draws) on out-of-fold predictions; decision-curve net benefit followed Vickers and Elkin~\cite{vickers_decision_2006} with 2{,}000-draw bootstrap bands.

Cox proportional hazards models~\cite{cox_regression_1972} using month 0 features served as comparators. We report Harrell's C~\cite{harrell__regression_2015} and inverse probability of censoring weighted (IPCW) C~\cite{uno_cstatistics_2011}. We also report AUC (95\% CI), calibration slope, and the Brier score~\cite{brier_verification_1950} for raw and calibrated probabilities.


\bgroup
\fixFloatSize{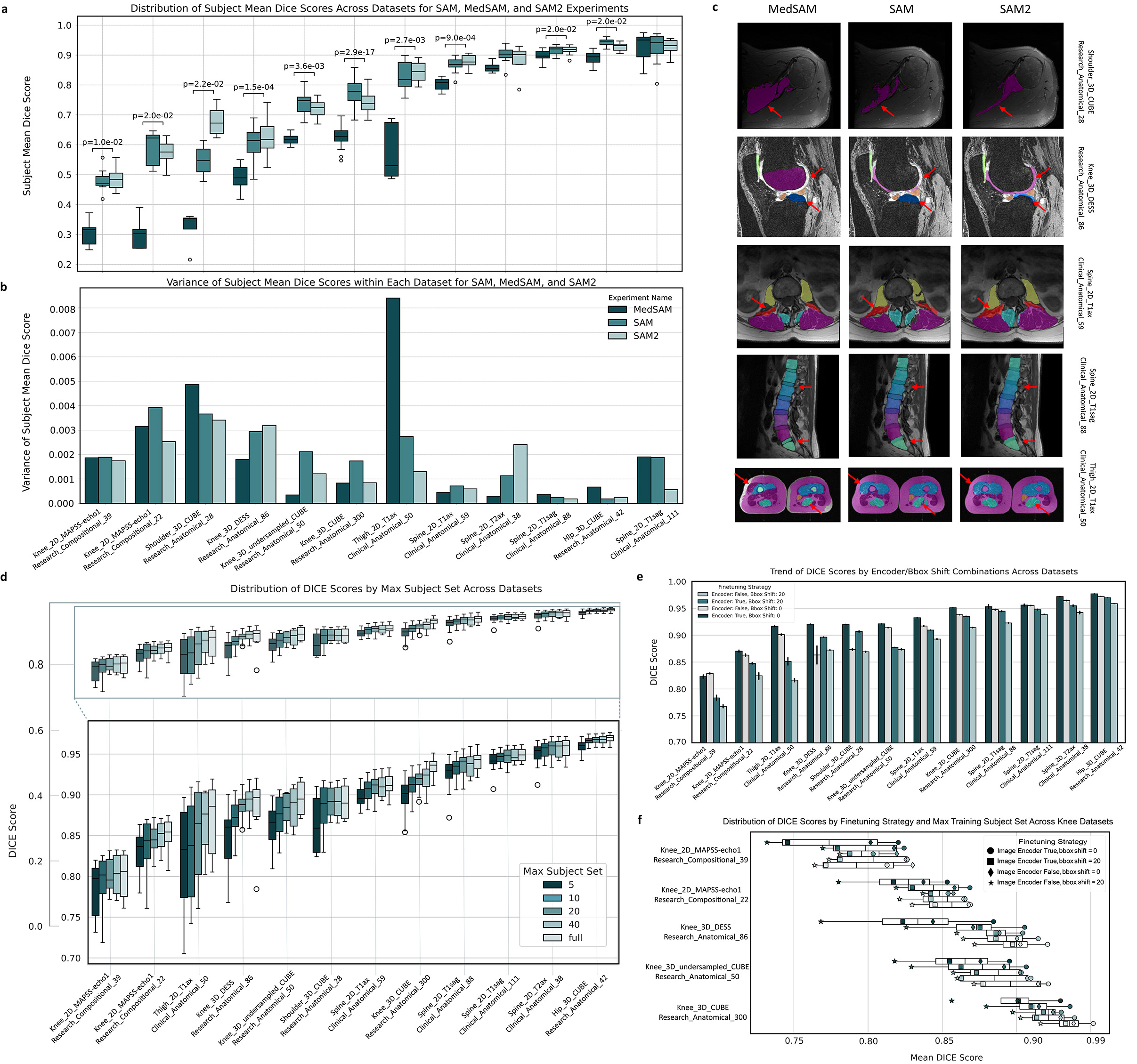}
\begin{figure*}[!b]

\captionsetup{position=above, singlelinecheck=false, justification=raggedright}
\centering
\caption[]{\textbf{Zero-shot and fine-tuned performance of SAM-family models across musculoskeletal MRI.}}
\makeatletter
\IfFileExists{images_final/main_fig_3_small.jpg}
  {\includegraphics[width=0.88\linewidth]{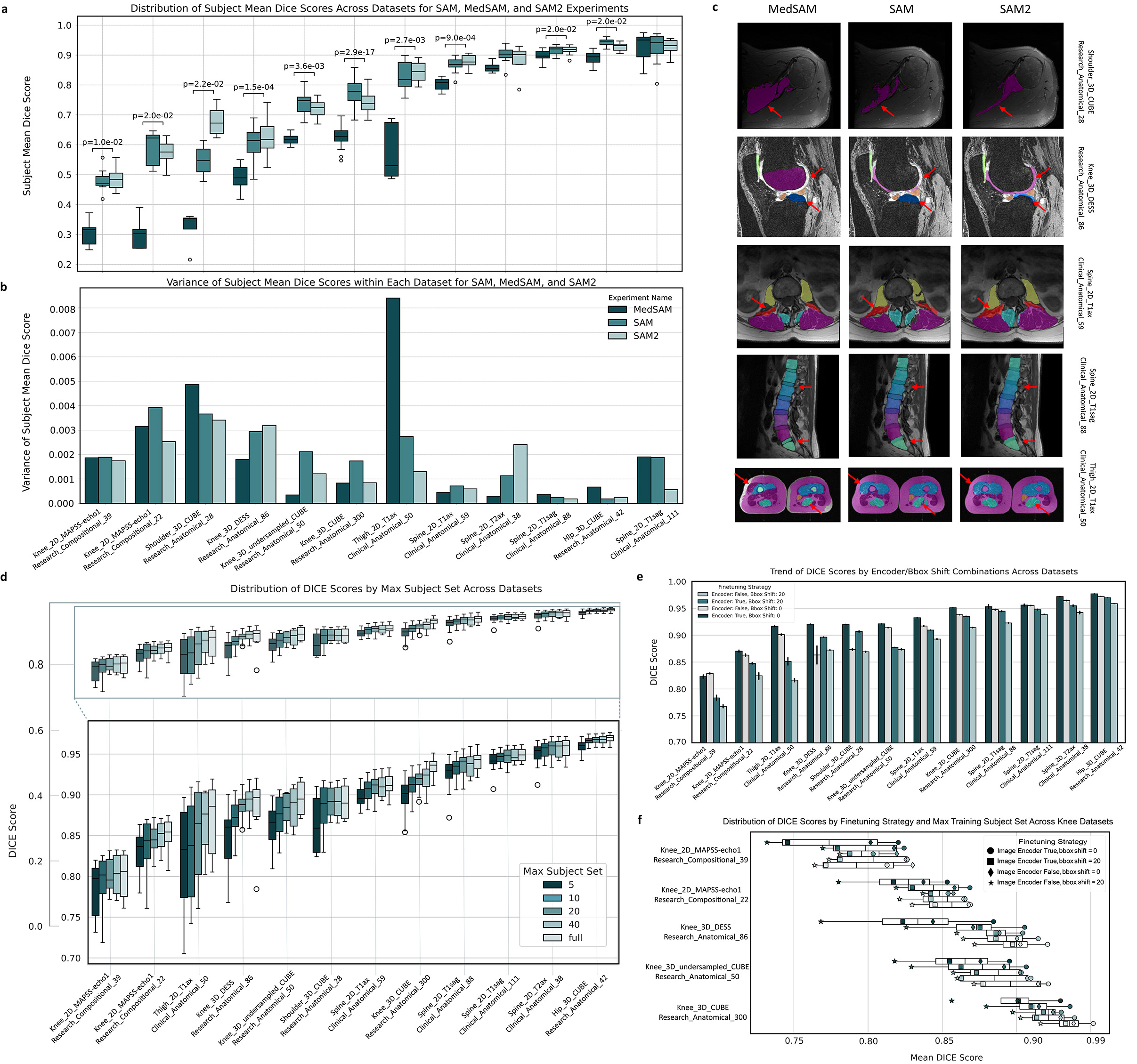}}
  {\includegraphics[width=0.88\linewidth]{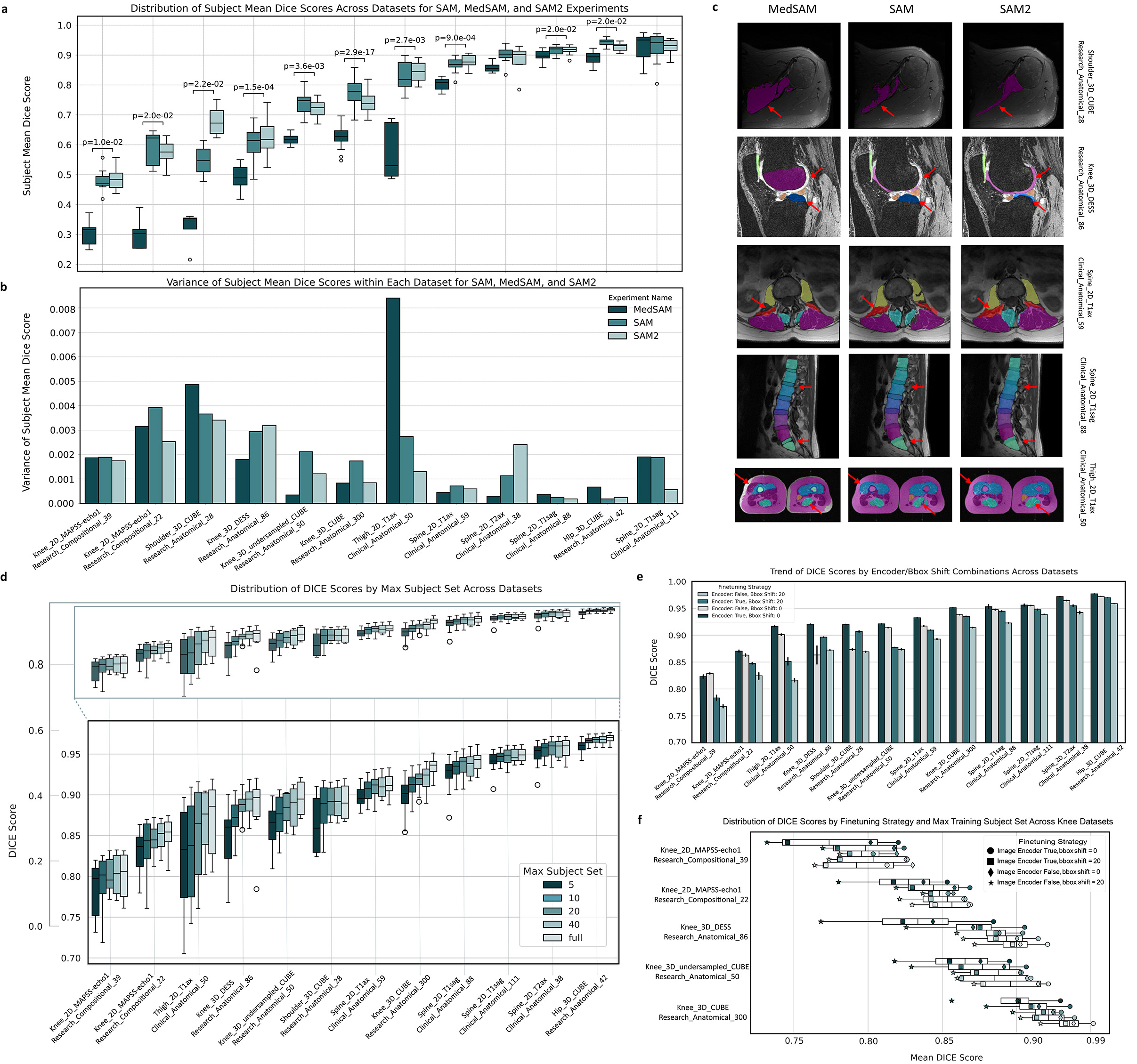}}
\makeatother

\captionsetup{position=below}

\begin{center}
\begin{minipage}{\linewidth}
\small\justify
\hyphenation{segmen-tation musculo-skeletal data-sets statis-tically signi-ficant}
\textbf{a)} Subject-level Dice across datasets for MedSAM, SAM, and SAM2. Brackets show Friedman test p values with Benjamini-Hochberg adjustment. Full statistics are in Supplementary Table S4. \textbf{b)} Variance of subject-level Dice within each dataset by model. \textbf{c)} Representative segmentations for five datasets. Columns show MedSAM, SAM, and SAM2. Rows show shoulder 3D CUBE, knee 3D DESS, spine 2D T1 axial, spine 2D T1 sagittal, and thigh 2D T1 axial. Arrows mark regions used for qualitative comparison. \textbf{d)} Subject-level Dice distributions by training size (5, 10, 20, 40, full) with an inset zoom near 0.9–1.0. \textbf{e)} Mean Dice with full-data fine-tuning under four strategies varying the image encoder (frozen vs fine-tuned) and bounding-box shift (0 vs 20 px). Error bars show inter-subject variability. \textbf{f)} Knee-only distributions across training sizes with points for the four strategies using the same marker shapes as in panel e.
\end{minipage}
\end{center}

\label{fig:baseline}
\end{figure*}
\egroup

\FloatBarrier


Demographics-only baselines use the same folds and calibration.

Sensitivity analyses repeated the pipeline with per-scan mean imputation for missing visits and with an extended 48$\rightarrow$120-month TKR horizon. Model explanation used Shapley additive explanations (SHAP): TreeSHAP~\cite{lundberg_local_2020} for random-forest TKR and SHAP's linear explainer~\cite{lundberg_unified_2017} for OA logistic regression on standardized inputs. Complete cohort definitions, threshold grids for decision curves, and sensitivity outputs appear in the Supplementary Information 5.10 (Tables \hyperref[fig:supp-s20]{S20}--S25; Figs. \hyperref[fig:supp-fig-s17]{S17}--S18).


\section{Results}

\subsection{Dataset Overview}

We analyzed 913 musculoskeletal MRI volumes (72,915 slices) spanning knee, hip, lumbar spine, shoulder, and thigh cohorts. The mix of high-resolution 3D (CUBE, DESS) and routine 2D spin-echo scans covers cartilage, bone, muscle, fat, and nerve labels (\hyperref[fig:supp-s0]{Table S0}). Participant age averaged 51 \ensuremath{\pm} 17 yr and weight 78 \ensuremath{\pm} 18 kg; sex distribution was balanced (461 women, 449 men) (\hyperref[fig:demographics]{Fig.~\ref*{fig:demographics}}; Supplementary Table S1). Imaging parameters (slice and pixel spacing, field strength, coil type) are summarized in Supplementary Table S2. \hyperref[fig:demographics]{Fig.~\ref*{fig:demographics}a–g} summarizes dataset composition, demographics, imaging parameters, label coverage, and representative slices.


\subsection{Foundation Models and Baseline (zero-shot) Assessment}

We benchmarked three Vision-Transformer segmenters: SAM (ViT-B, 91 M parameters), SAM2 (hiera\_base\_plus), and MedSAM (ViT-B pre-trained on medical data), using bounding-box prompts from manual masks. We compared model rankings using a Friedman test, followed by Wilcoxon signed-rank tests with Benjamini-Hochberg control at 5\%; full statistics are in Supplementary Tables S3–S5. All tests were two-sided with \(\alpha=0.05\), and we report test statistics, unadjusted, and FDR-adjusted \(p\)-values.

Models demonstrated statistically significant differences in baseline performance (Friedman $q<0.05$), with SAM2 consistently outperforming both SAM and MedSAM on complex anatomical structures \hyperref[fig:baseline]{Fig.~\ref*{fig:baseline}a,b}. SAM2 led on complex anatomy, such as shoulder CUBE Dice 0.68 versus 0.55 (SAM) and 0.32 (MedSAM), and thigh axial Dice 0.84 versus 0.83 and 0.58 (all $p<0.01$). On simpler spine T1 sagittal scans, differences were non-significant. Representative overlays can be seen in \hyperref[fig:baseline]{Fig.~\ref*{fig:baseline}c}; full per-dataset statistics are in Supplementary Tables S3–S5.


\subsection{Fine-tuning and Data Efficiency}
Fine-tuning raised Dice across all datasets (median +0.14). The 25th--75th percentile shifted from 0.59--0.89 to 0.92--0.96, and 83\% of fine-tuned scores reached $\geq$0.90 (\hyperref[table:tiered_table]{Table~\ref*{table:tiered_table}}). Median gains were +0.30 for MedSAM, +0.14 for SAM2, and +0.12 for SAM, consistent with their baseline gaps. In \textit{Knee\_3D\_DESS}, for example, SAM2 climbed from 0.62 to 0.94 after full encoder--decoder training. Updating both encoder and decoder without prompt shift provided uniform benefits across anatomies. Subject-level distributions across training tiers are shown in \hyperref[fig:baseline]{Fig.~\ref*{fig:baseline}d}; scenario summaries and a knee-specific breakdown appear in \hyperref[fig:baseline]{Fig.~\ref*{fig:baseline}e-f}. Scenario details appear in Supplementary \hyperref[fig:supp-s6]{Table~S6}; top-performing model-strategy pairs and tissue-level results are listed in Supplementary Tables S7--S9, with Dice distributions in Supplementary Fig.~S1.

In addition to mean Dice (\hyperref[table:tiered_table]{Table~\ref*{table:tiered_table}}), we summarize subject-level lower-tail performance (median, 5th percentile, and minimum Dice) to quantify worst-case behavior; compact summaries aligned to Table~1 are provided in Supplementary Data D56--D57, with the full experiment set in Supplementary Data D55.

The largest determinant of performance was whether the backbone was fine-tuned for the target label set, with full encoder--decoder fine-tuning consistently shifting Dice into the 0.92--0.96 range across anatomies (\hyperref[fig:baseline]{Fig.~\ref*{fig:baseline}}; \hyperref[table:tiered_table]{Table~\ref*{table:tiered_table}}). In the zero-shot setting, SAM2 tended to be the most robust across anatomies and low-contrast structures. MedSAM showed the largest relative gains after fine-tuning, consistent with medical-domain pretraining, while SAM remained a strong baseline when prioritizing broad compatibility and minimal configuration. In practice, these results support using a pooled fine-tuned model (e.g., mskSAM2) as a strong default. Anatomy- or dataset-specific fine-tuning can be applied when targeting thin or low-contrast structures or when protocol shift is expected.


\subsection{Performance by Tissue and Protocol}
We examined whether fine-tuning benefits extended across diverse tissue types and MRI protocols, as clinical deployment requires consistent performance regardless of acquisition parameters. Fine-tuned models achieved consistent gains across tissue classes and MRI protocols. For \textit{Knee\_3D\_DESS}, mskSAM2 exceeded 0.93 for every cartilage and meniscus label (mean 0.94); zero-shot SAM and SAM2 remained near 0.62 and MedSAM at 0.49, with baseline menisci below 0.55 (Supplementary Fig.~S2). In \textit{Spine\_2D\_T2ax}, the same model reached Dice 0.978 for cartilage, 0.965 for bone, and 0.969 for the spinal nerve bundle, an absolute gain of 18 points over zero-shot SAM2 for the nerve class, which is important for decompression planning (Supplementary Fig.~S3).

Fine-tuning also helped in low-contrast, adipose-rich regions. In \textit{Thigh\_2D\_T1ax}, Dice rose from 0.76 to 0.85 for subcutaneous fat and from 0.82 to 0.89 for femoral cortex; muscle compartments passed 0.95, and a muscle-focused mixed\_Muscle\_SAM averaged 0.92 (Supplementary Fig.~S4). An 8$\times$ undersampled knee CUBE sequence, withheld during training, probed robustness to low signal-to-noise ratio (SNR): baseline models lost 4--5 Dice points on cartilage, yet fine-tuned mskSAM2 fell by only 0.05 (0.92 to 0.87) and kept bone Dice 0.97.

Complete anatomy-, tissue-, and strategy-stratified results appear in Supplementary Figs.~S2--S4, \hyperref[table:tiered_table]{Table \ref*{table:tiered_table}}, and Supplementary Data Tables D1--D24. These findings suggest that encoder--decoder fine-tuning may be necessary for curved or low-contrast tissues. They also show that single global Dice values can unintentionally conceal clinically important disparities.

Lower-tail cartilage performance improved substantially with fine-tuning: across knee cartilage labels, the 5th percentile Dice ranged from $\sim$0.80--0.92 for the fine-tuned models (minimum $\sim$0.78--0.91), whereas zero-shot cartilage Dice could fall as low as $\sim$0.25 on the smallest MAPSS cohorts (Supplementary Data D56).

These tissue- and protocol-specific improvements demonstrate that foundation models can achieve clinical-grade performance across the heterogeneous imaging conditions typical of MSK radiology services.

\subsection{Influence of MRI Acquisition Parameters}
Mixed-effects modeling (\hyperref[fig:biomarkers]{Fig.~\ref*{fig:biomarkers}a}) showed that higher flip angle and finer pixel spacing were associated with improved Dice, whereas longer echo time showed a modest negative association. These associations are plausible because flip angle and echo time influence tissue contrast and 


\bgroup
\fixFloatSize{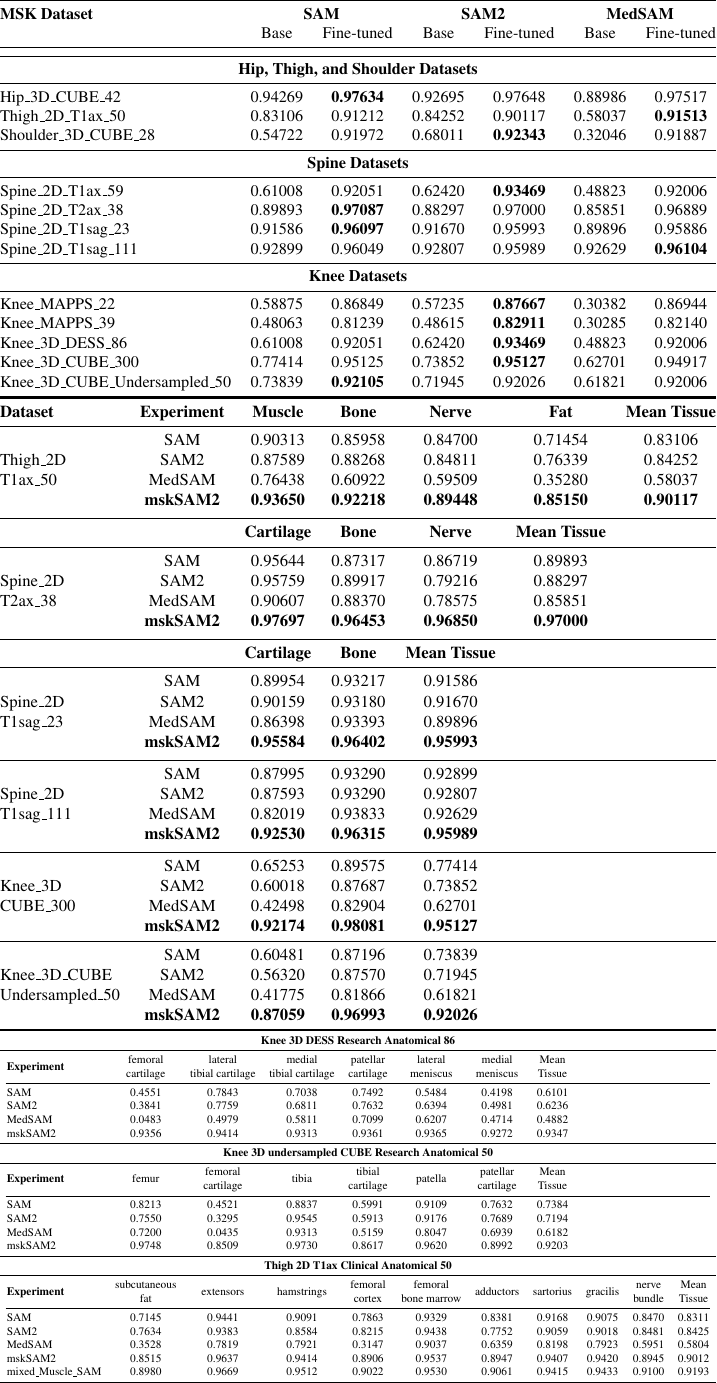}
\begin{table}[H]

\captionsetup{position=above, singlelinecheck=false, justification=raggedright}
\centering
\caption[]{\textbf{Tiered evaluation of SAM variants on musculoskeletal MRI.}}

\makeatletter\IfFileExists{images_final/Table_1.pdf}{\includegraphics[width=1.0\linewidth]{images_final/Table_1.pdf}}{\includegraphics[width=1.0\linewidth]{Table_1.pdf}}
\makeatother 

\captionsetup{position=below,font=scriptsize, skip=4pt}
\begin{center}
\begin{minipage}{\linewidth}
\small\justify

Dice scores are grouped into three blocks. \textbf{1.} Dataset level (see previous figure). \textbf{2.} Tissue level lists cartilage, bone, muscle, nerve, and fat for datasets that include multi-tissue labels. \textbf{3.} Label level breaks those tissues into individual structures: knee rows cover six cartilage compartments, three subchondral bones, and both menisci; spine, hip, and shoulder follow the same principle. The thigh panel adds quadriceps, hamstrings, adductors, etc., plus a muscle-focused model (\textit{mixed\_Muscle\_SAM}) that shows the benefit of task-specific fine-tuning. Columns compare baseline SAM, SAM2, and MedSAM with their full-dataset fine-tuned counterparts (\textit{mskSAM, mskSAM2}). Fine-tuning updated both encoder and decoder without bounding-box shift augmentation. This tiered layout captures differences in model behavior across datasets, tissue categories, and anatomical detail, allowing closer inspection of segmentation performance at increasing levels of complexity. 
\end{minipage}
\end{center}

\label{table:tiered_table}

\end{table}
\egroup



\bgroup
\begin{figure*}[!htbp]

\captionsetup{position=above, singlelinecheck=false, justification=raggedright}
\centering
\caption[]{\textbf{Analysis of MRI acquisition parameters and segmentation agreement for fine-tuned SAM models in musculoskeletal MRI.}}

\makeatletter\IfFileExists{images_final/main_fig_4_small.png}{\includegraphics[width=1.0\linewidth]{images_final/main_fig_4_small.png}}{\includegraphics[width=1.0\linewidth]{main_fig_4_small.png}}
\makeatother 

\captionsetup{position=below}

\begin{center}
\begin{minipage}{\linewidth}
\small\justify
\textbf{a)} Fixed effects estimates with 95\% confidence intervals for three fine-tuning experiment types: (1) single-dataset training, (2) grouped training based on anatomical region, tissue type, or MRI sequence, and (3) a full-dataset experiment incorporating all musculoskeletal MRI data (mskSAM). The vertical axis lists MRI acquisition parameters, including slice thickness, image dimensionality (2D vs. 3D), flip angle, TE, and pixel spacing. Each point reflects the estimated effect size of a given parameter on segmentation performance, with models fit using SAM weights as the reference. All available datasets and model variants (SAM, SAM2, MedSAM) are included in the analysis. \textbf{b)} Concordance between automated and manual segmentation measurements for six biomarker types: muscle volume, disc height, cartilage thickness, T\ensuremath{_{1\ensuremath{\rho}}}, T\ensuremath{_{2}}, and bone volume. Each row corresponds to one biomarker. Bland-Altman plots (left column) compare value distributions, and regression plots (right column) assess correlation. Linear regression metrics (R\ensuremath{^2}, p-value, ICC3) are used for normally distributed data, while Gaussian process models with Spearman's \ensuremath{\rho}, p-values, and bootstrapped ICCs are used for non-normal distributions. See Supplementary Information §5.7; Supplementary Tables \hyperref[fig:supp-s13]{S13}--S16; and Supplementary Figs. S5–S10 for statistical tests and per-dataset plots.
\end{minipage}
\end{center}

\label{fig:biomarkers}
\end{figure*}
\egroup

\FloatBarrier
\clearpage


signal-to-noise ratio, which in turn affect boundary visibility for cartilage, muscle, and discs. Longer TE and lower flip angle can reduce effective contrast or introduce additional blurring/partial-volume effects in some sequences, making thin structures harder to delineate.

We note that protocol parameters can also be partially confounded with dataset-specific factors (anatomy, label set granularity, and task difficulty); accordingly, we interpret parameter effects descriptively and report full model specifications and uncertainty estimates in Supplementary Information 5.6.

These effects were attenuated after full multi-dataset fine-tuning. Complete coefficients, interaction terms, preprocessing steps, and model checks are detailed in Supplementary Materials 7.1 (Tables~S10--S12 and Data Tables~D25--D26).


\subsection{Biomarker Agreement with Expert Measurements}
Fine-tuned segmentations agreed with experts across all six biomarkers. Limits of agreement and regression plots appear in \hyperref[fig:biomarkers]{Fig.~\ref*{fig:biomarkers}b}, with dataset-specific views in Supplementary Figs.~S5--S10. Normality (Shapiro-Wilk) and variance (Levene) checks (Supplementary \hyperref[fig:supp-s13]{Table~S13}) dictated whether parametric ICC(3,1) or a bootstrap mixed-effects ICC was applied, and whether agreement was summarized by the mean or the median value for each metric.

For cartilage thickness, the lateral tibial compartment reached ICC 0.996 on knee DESS and remained $\geq$0.89 on the lower-contrast CUBE protocol; Bland-Altman limits were $\pm$0.18\,mm, comparable to published test--retest precision for MRI-based cartilage morphometry \cite{eckstein_quantitative_2011}.

Disc-height ICC at L1--L2 was 0.97 with bias $<$0.6\,mm. Quadriceps-volume ICC was 0.99 in thigh MRI, and total lumbar muscle volume achieved ICC 1.00, both tracking manual estimates within 2\%.

Relaxometry results indicate fidelity for biochemical imaging. Across six cartilage regions in knee MAPSS datasets, median mixed-effects ICCs were 0.97 for $T_{1\rho}$ and 0.89 for $T_{2}$; bias stayed below 1.5\,ms. This veracity for both structural and compositional biomarkers demonstrates that our pipeline can support unified measurement extraction for musculoskeletal research and clinical applications.


\bgroup
\fixFloatSize{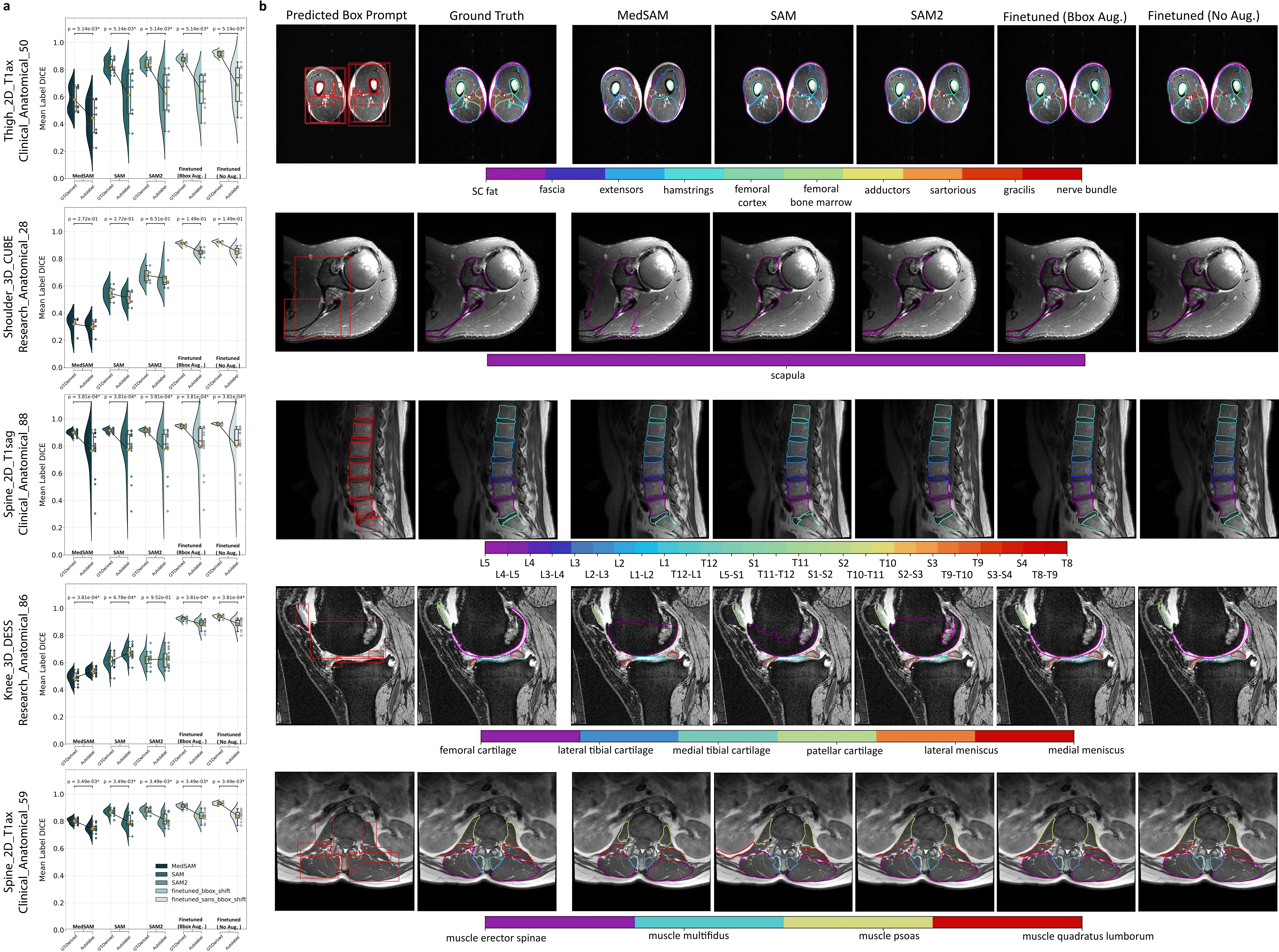}
\begin{figure*}[!htbp]

\captionsetup{position=above, singlelinecheck=false, justification=raggedright}
\centering
\caption[]{\textbf{Comparative evaluation of segmentation models using ground truth and automated bounding box prompts across musculoskeletal MRI datasets.}}

\makeatletter\IfFileExists{images_final/main_fig_5_tiny.png}{\includegraphics{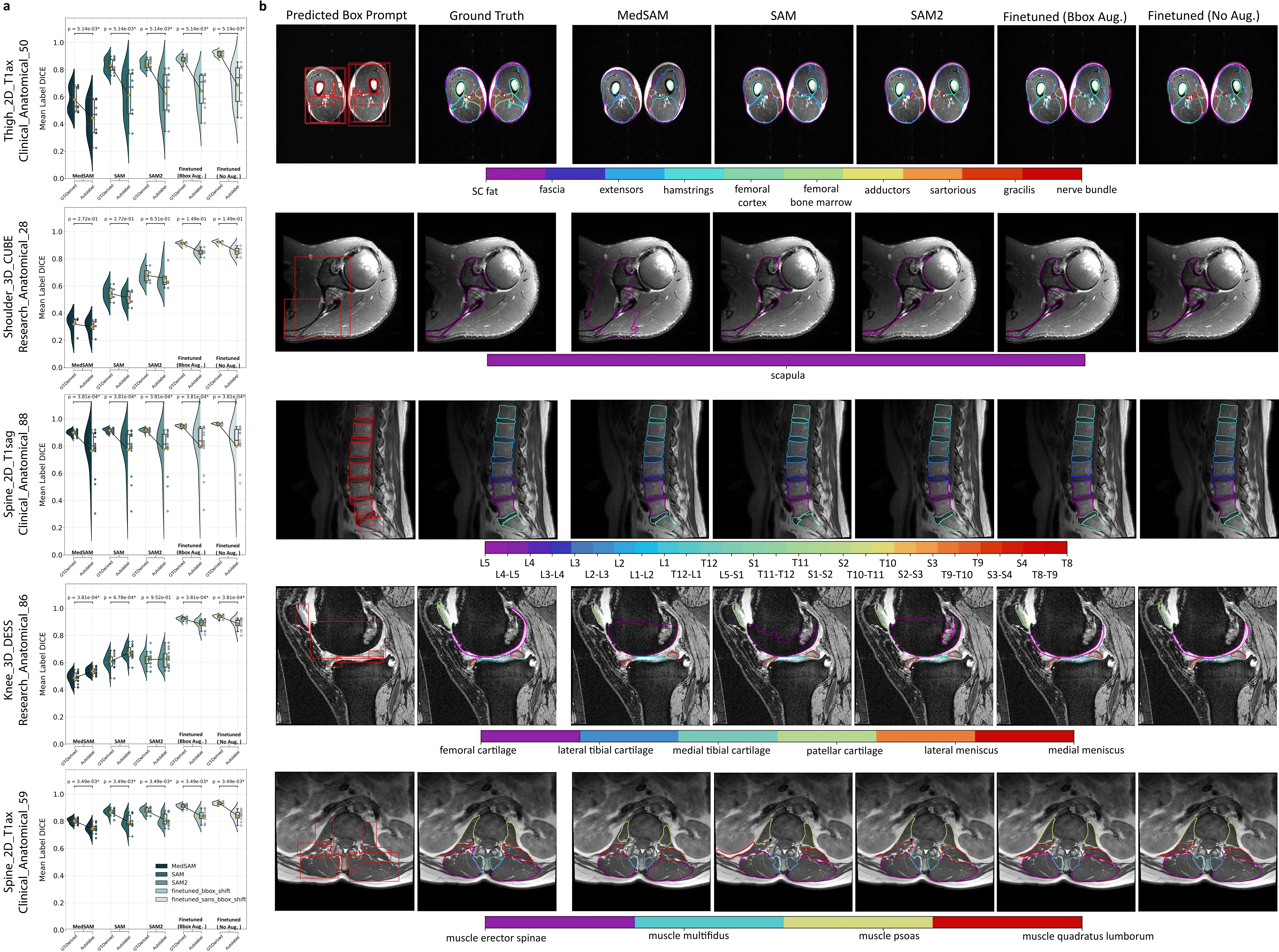}}{\includegraphics{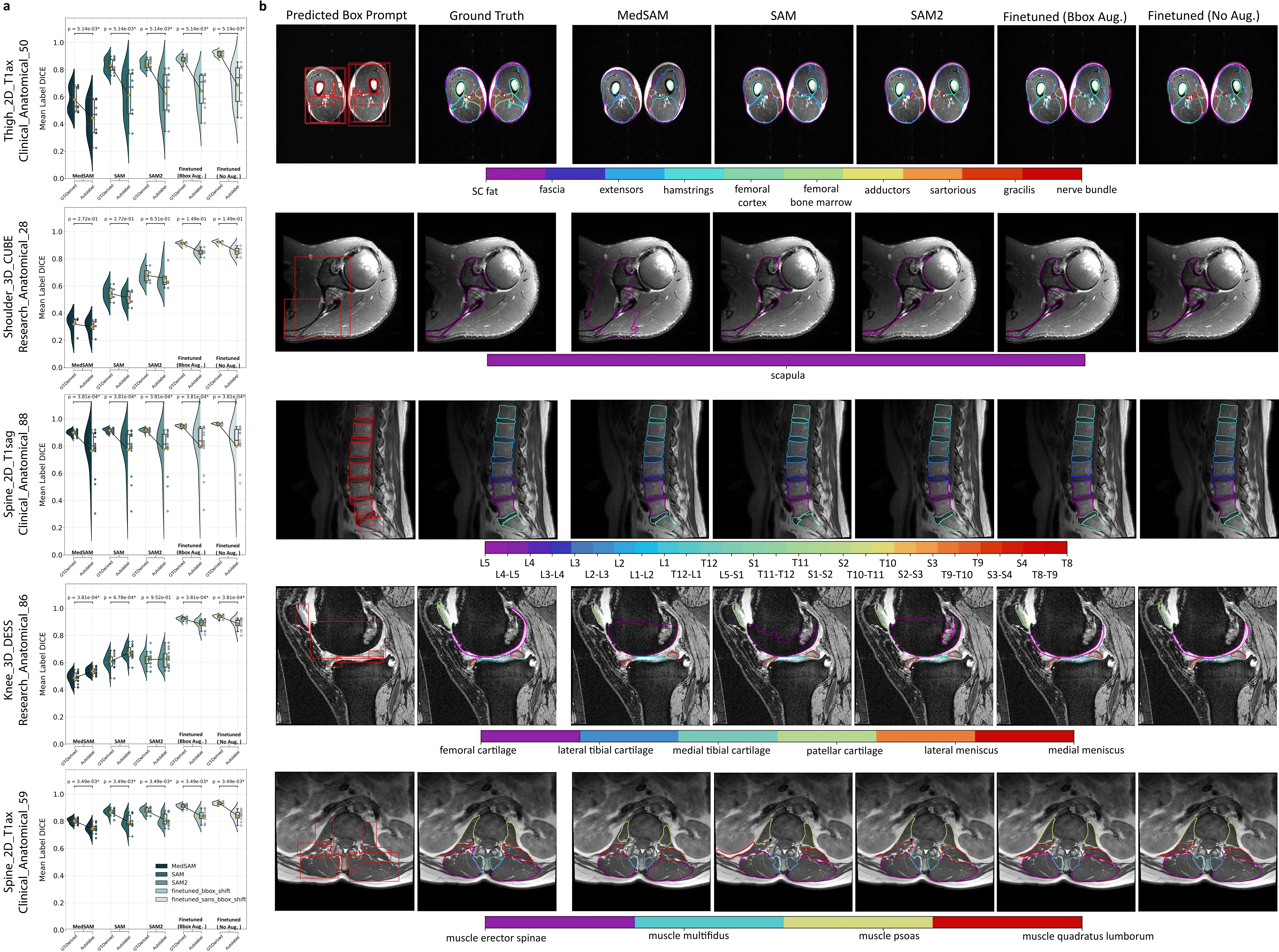}}
\makeatother 

\captionsetup{position=below}
\begin{center}
\begin{minipage}{\linewidth}
\small\justify
\textbf{a)} Raincloud plots comparing subject-level Dice scores from five segmentation models: MedSAM, SAM, SAM2, Fine-tuned (with bounding box augmentation), and Fine-tuned (without augmentation). Each raincloud plot combines a violin plot (distribution), a strip plot (individual subject scores), and a box plot (interquartile range and median). For each model, paired plots display performance using ground truth-derived prompts (left) and automated detection-based prompts (right, AutoLabel). Statistical differences between prompt conditions are assessed using Wilcoxon rank-sum tests, with p-values adjusted for multiple comparisons using the Benjamini-Hochberg correction. Adjusted p-values are reported above each model pair. \textbf{b)} Representative segmentations from five MRI datasets (top to bottom): \textit{Thigh\_2D\_T1ax\_Clinical\_Anatomical\_50}, \textit{Shoulder\_3D\_CUBE\_Research\_Anatomical\_28}, \textit{Spine\_2D\_T1sag\_Clinical\_Anatomical\_88}, \textit{Knee\_3D\_DESS\_Research\_Anatomical\_86}, and \textit{Spine\_2D\_T1ax\_Clinical\_Anatomical\_59}. Each row shows the original MRI slice with predicted bounding boxes, followed by ground truth overlay, model prediction overlay, ground truth contours, and predicted contours. These examples show how the source of the bounding box prompt can influence segmentation accuracy across different anatomical targets and imaging protocols. Statistics and examples: Supplementary Table S17; Supplementary Figs. S11–S15.
\end{minipage}
\end{center}

\label{fig:autolabel}
\end{figure*}
\egroup



\subsection{AutoLabel Prompt Sensitivity}

Replacing manual boxes with YOLO-generated prompts lowered mean Dice by 0.05–0.11 across five datasets (\hyperref[fig:autolabel]{Fig.~\ref*{fig:autolabel}a}). Losses were smallest on \textit{Shoulder\_3D\_CUBE} and largest on \textit{Thigh\_2D\_T1ax}. Pairwise Wilcoxon tests confirmed differences for most model and prompt pairs (Supplementary Table S17). Qualitative examples appear in \hyperref[fig:autolabel]{Fig.~\ref*{fig:autolabel}b} and Supplementary Figs. S11–S15. Methods and training details are in Supplementary Materials 6.4.1–6.4.2; post-processing/evaluation workflow in 5.8.1.

For throughput, AutoLabel processed a 3D knee in a median 30.1\,s on an A100 40\,GB and 42.6\,s on a TITAN RTX, excluding file I/O and end-run aggregation (30 scans each). On TITAN, medians were 8.3\,s for detection, 21.9\,s for segmentation, and 12.4\,s for per-slice biomarker updates; timing methods are in Supplementary Materials 5.9.
Having established measurement reliability across diverse conditions, we evaluated two clinical applications of the validated pipeline.


\bgroup

\begin{figure*}[!t]

\captionsetup{position=above, singlelinecheck=false, justification=raggedright}
\centering
\caption[]{\textbf{Automated knee MRI triage and case routing.}}

\makeatletter\IfFileExists{images_final/main_fig_6.pdf}{\includegraphics[width=0.8\linewidth]{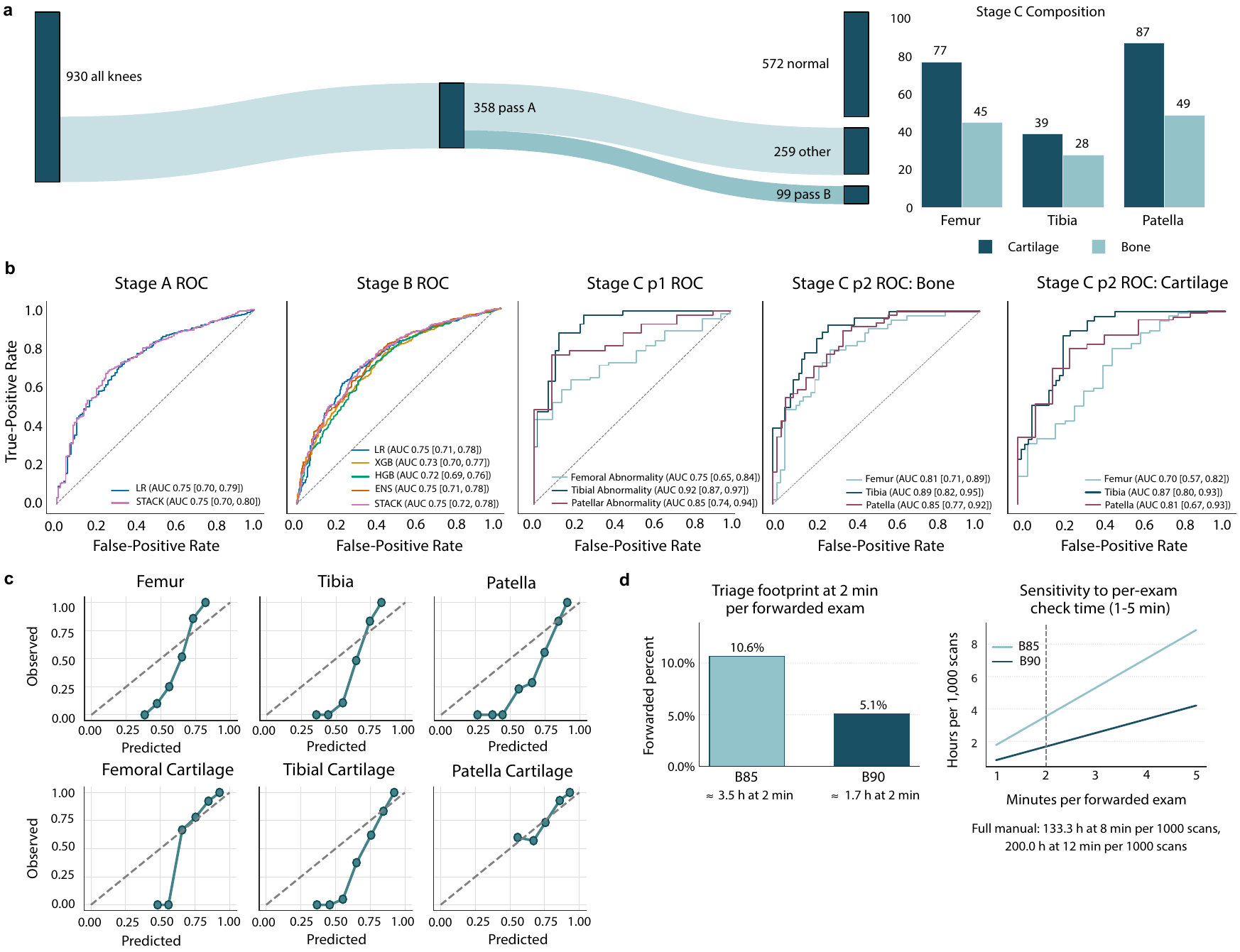}}{\includegraphics[width=0.8\linewidth]{main_fig_6.pdf}}
\makeatother 

\captionsetup{position=below}

\begin{center}
\begin{minipage}{\linewidth}
\small\justify
\textbf{a)} Cohort flow with preset operating points. Left, Sankey diagram from all knees to Stage A and Stage B. Right, counts by joint and tissue within Stage-C evaluation. \textbf{b)} ROC curves for Stage A, Stage B, Stage C part 1 (joint-level abnormality), and Stage C part 2 (tissue-specific tasks). Markers indicate the pre-specified specificity targets used for routing. \textbf{c)} Calibration at the a90p\_b85p operating point for the six Stage-C tasks. Points show observed event frequency vs predicted probability in equal-width bins; the gray dashed line indicates perfect calibration. No probability recalibration was applied. \textbf{d)} Workload summary for triage verification. Left, forwarded percent at the two Stage-B settings with hours per 1,000 scans assuming a 2-minute triage check for forwarded exams. Right, sensitivity of hours per 1,000 scans to per-exam verification time from 1 to 5 minutes, with a marker at 2 minutes. Full manual review at 8 and 12 minutes per exam is shown as a context baseline. This panel quantifies triage verification only. All exams still receive full diagnostic review. See Supplementary Information §5.9–§5.9.1; Supplementary Tables \hyperref[fig:supp-s18]{S18}--19; and Supplementary Fig. S16.
\end{minipage}
\end{center}

\label{fig:clinical}
\end{figure*}
\egroup



\subsection{Triage performance}

We evaluated whether automated biomarkers could safely reduce radiologist workload while maintaining sensitivity for clinically significant pathology. \hyperref[fig:clinical]{Fig.~\ref*{fig:clinical}a} shows the cohort flow and routed composition. Applying the AutoLabel $\rightarrow$ biomarker $\rightarrow$ classifier cascade to 930 knee scans yielded the operating characteristics in \hyperref[fig:clinical]{Fig.~\ref*{fig:clinical}b} and Supplementary \hyperref[fig:supp-s18]{Table~S18}; calibration curves are shown in \hyperref[fig:clinical]{Fig.~\ref*{fig:clinical}c} and workload estimates in \hyperref[fig:clinical]{Fig.~\ref*{fig:clinical}d}. Full demographic profiles, stage-wise performance, and workload estimates are provided in Supplementary Materials~7.2. 

Stage A at 90\% specificity removed 62\% of scans (AUC 0.75). At the Stage B thresholds, the cascade forwarded 99 of 930 knees (10.6\%) at B85 and 47 of 930 (5.1\%) at B90. If a 2\,min check is applied to forwarded studies only, the triage step would require approximately 3.5\,h per 1,000 scans at B85, and 1.7\,h per 1,000 at B90 (\hyperref[fig:clinical]{Fig.~\ref*{fig:clinical}d}). Calibration curves are shown in Supplementary Fig.~S16. Confusion counts and tissue-level tables appear in Supplementary Table~S19. Threshold definitions are reported in Supplementary Tables \hyperref[fig:supp-s18]{S18}--S19; workload math is in Supplementary Materials 5.9.1.


\subsection{Prediction of Knee Replacement and Incident Osteoarthritis}

At the 48-month landmark (\hyperref[fig:landmark]{Fig.~\ref*{fig:landmark}a}), we screened learning algorithms (\hyperref[fig:landmark]{Fig.~\ref*{fig:landmark}b}) and selected final models (\hyperref[fig:landmark]{Fig.~\ref*{fig:landmark}c}); explanations and clinical utility are summarized in \hyperref[fig:landmark]{Fig.~\ref*{fig:landmark}d,e}. 

Using biomarkers through 48 months in 994 knees at risk, the TKR model reached AUC 0.76 (95\% CI 0.72–0.80) (\hyperref[fig:landmark]{Fig.~\ref*{fig:landmark}c}), calibration slope 0.997, and Brier 0.099, with net benefit across 11–50\% risk (\hyperref[fig:landmark]{Fig.~\ref*{fig:landmark}e}). Full model outputs, extended horizon results, and performance metrics are presented in Supplementary Materials~7.3. A demographics-only baseline at the same landmark achieved AUC 0.60 (0.55–0.64). SHAP ranked lateral tibial cartilage features highest, followed by patellar cartilage; the lateral meniscus also contributed (\hyperref[fig:landmark]{Fig.~\ref*{fig:landmark}d}).


\bgroup
\begin{figure*}[!htbp]

\captionsetup{position=above, singlelinecheck=false, justification=raggedright}
\centering
\caption[]{\textbf{Landmark risk models from automated knee MRI biomarkers.}}

\makeatletter\IfFileExists{images_final/main_fig_7.pdf}{\includegraphics[width=1.0\linewidth]{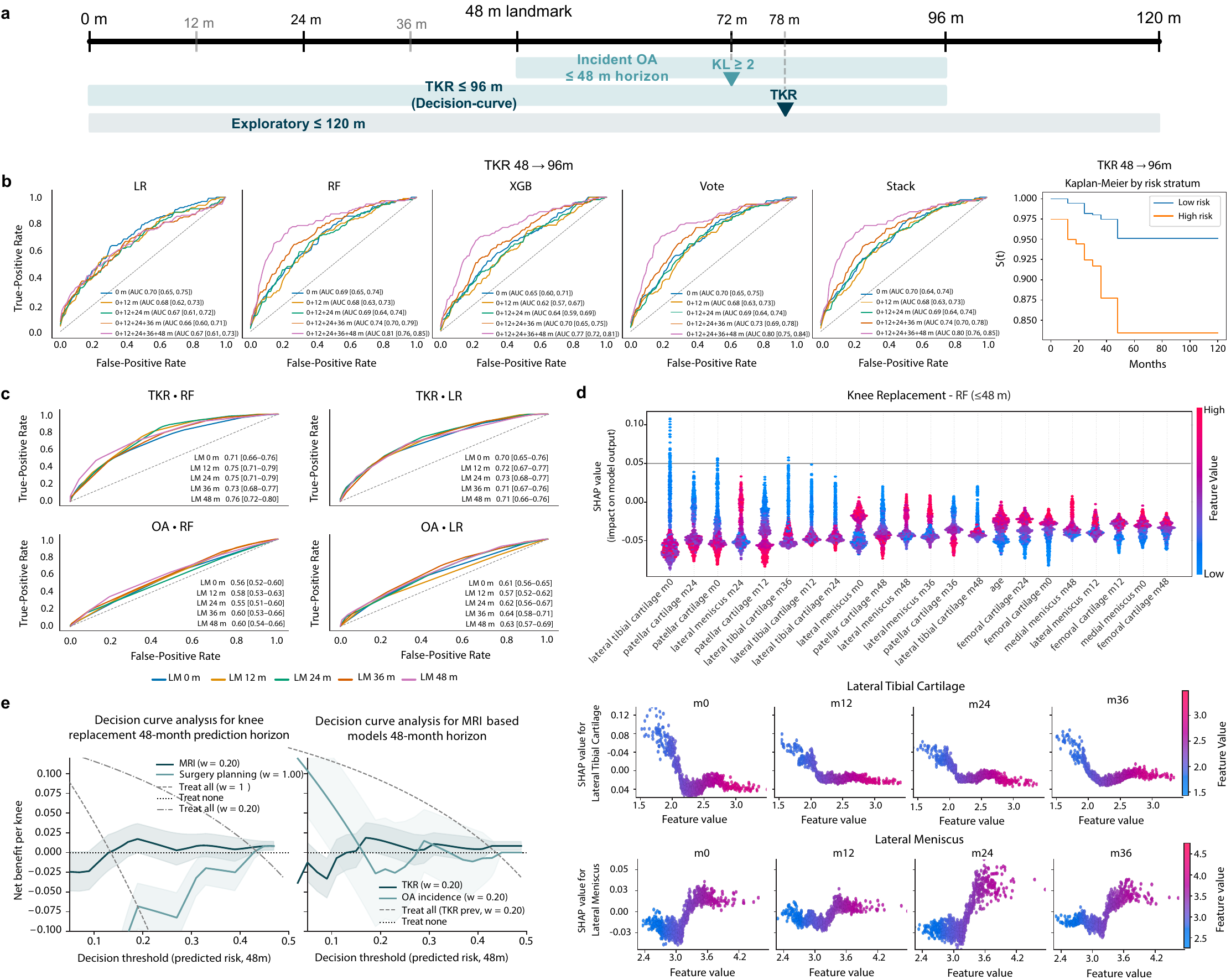}}{\includegraphics[width=1.0\linewidth]{main_fig_7.pdf}}
\makeatother 

\captionsetup{position=below, font=scriptsize, skip=4pt}
\begin{center}
\begin{minipage}{\linewidth}

\small\justify
\textbf{a)} Landmark design at 48 months. Main horizon: TKR over 48$\rightarrow$96 months. Exploratory horizon: 48$\rightarrow$120 months. OA window: \ensuremath{\leq}48 months. \mbox{}\newline \textbf{b)} Architecture screen for TKR at the 48-month landmark. ROC curves for LR, RF, XGB, soft-vote, and stack are shown for screening only. The rightmost panel shows a Kaplan-Meier curve for TKR 48$\rightarrow$96 months using the RF screen model with a median split. These panels are used to choose an algorithm; all downstream analyses use the final models in panel c. \mbox{}\newline \textbf{c)} Final landmark models used for the study. Left: TKR 48$\rightarrow$96 months, RF and LR overlaid. Right: OA incidence \ensuremath{\leq}48 months, LR and RF overlaid. Legends report AUC with 95\% CIs. \mbox{}\newline \textbf{d)} SHAP summary for the final TKR model at the 48-month landmark. Beeswarm for the top 20 features; positive SHAP values increase predicted risk. \mbox{}\newline \textbf{e)} Left: decision-curve analysis at the 48-month landmark. TKR under standard, MRI 0.20, and surgery 1.00 penalties, and a panel comparing TKR vs OA under the MRI-triage penalty. Shaded bands show 95\% bootstrap. Right: SHAP dependence examples for lateral tibial cartilage and lateral meniscus thickness at baseline, 12, 24, and 36 months; color encodes the raw feature value. OA SHAP and additional dependence plots are provided in Supplementary Information §5.10; Supplementary Tables \hyperref[fig:supp-s20]{S20}--S25; and Supplementary Figs. \hyperref[fig:supp-fig-s17]{S17}--S18. \mbox{}\newline\newline Note: panel b shows the screening step only; survival, SHAP, and decision-curve results use the final landmark models in panel c. \mbox{}\newline\newline Abbreviations: TKR, total knee replacement; OA, osteoarthritis; ROC, receiver operating characteristic; LR, logistic regression; RF, random forest; XGB, XGBoost; CI, confidence interval.

\end{minipage}
\end{center}

\label{fig:landmark}
\end{figure*}
\egroup

\FloatBarrier
\clearpage


For incident radiographic osteoarthritis, logistic regression reached AUC 0.63 (0.57–0.69) (Supplementary Table~S22). Extended horizons, full metrics, and sensitivity analyses appear in Supplementary Tables \hyperref[fig:supp-s20]{S20}--S25, with curves in \hyperref[fig:supp-fig-s17]{Fig.~S17} and SHAP plots in Fig.~S18. Main results are summarized in \hyperref[fig:landmark]{Fig.~\ref*{fig:landmark}}. Cohort, features, modeling, and calibration are detailed in Supplementary Information~5.10.1–5.10.4; SHAP details in 5.10.6.


\section{Discussion}

Our study shows that foundation model segmentation can move from technical capability to clinical utility when biomarker fidelity guides system design in musculoskeletal imaging. We achieved three key deliverables: validated segmentation across diverse anatomies, reproducible quantitative biomarkers verified against expert measurements (ICCs up to 0.99), and clinical decision support tools, which address both operational efficiency and patient-specific risk prediction. The model-agnostic architecture supports sustainable deployment because new segmentation backbones can be integrated without affecting biomarker extraction or decision layers.

To establish clinical trust, we prioritized clinically relevant biomarker fidelity over segmentation metrics alone. Post-fine-tuning measurements showed agreement with expert references within clinically relevant tolerance ranges across all tested tissues and protocols. For cartilage thickness, Bland-Altman limits of \ensuremath{\pm}0.18\,mm were comparable to published test--retest precision and smallest-detectable-change estimates for MRI-based cartilage morphometry \cite{eckstein_quantitative_2011}. Our mixed-effects analysis indicated that imaging parameters affected segmentation accuracy, though multi-protocol training appeared to reduce these effects. This finding supports the potential value of unified models over protocol-specific approaches. At the parameter level, higher flip angle was associated with improved Dice, whereas longer echo time (TE) was associated with decreased Dice. These settings plausibly alter contrast-to-noise and boundary conspicuity; longer TE can also reduce SNR through T2 decay and blur tissue interfaces relevant for segmentation. This measurement reliability enabled our two proof-of-concept applications to demonstrate precision medicine at different timescales.

AutoLabel automation proved essential for scalability. While detector-generated prompts reduced Dice by 0.05--0.11 compared to manual boxes, this trade-off enables population-scale deployment without annotation bottlenecks. Automated prompts showed consistent performance when paired with validated segmentation, which could facilitate standardized measurements across sites. These technical foundations (high-quality segmentation, clinically sound biomarkers, and scalable automation) demonstrate that system readiness for precision MSK imaging emerges from biomarker-centric design rather than segmentation scores alone.

In this framework, imaging-derived measurements can support near-term prioritization and longer-horizon risk stratification. We evaluated two downstream workflows built on the same segmentation and feature extraction module. The first was a triage cascade that computed a small set of cartilage and bone biomarkers per knee, applied adjustable thresholds, and routed only flagged studies for detailed review. Threshold tuning managed caseload while maintaining sensitivity. This provisional tool design and evaluation suggests that reliance on standardized measurements rather than site-specific heuristics may foster portability across scanners and protocols. Current rules focus on cartilage and bone; incorporating meniscus and ligament metrics could improve coverage for knees with isolated soft-tissue injury without slowing throughput.

In the intended workflow, AutoLabel can run after acquisition to generate segmentations, overlay thumbnails for plausibility checks. The system then produces a structured biomarker report. This compact, standardized report can also serve as a longitudinal measurement record, enabling within-patient tracking and cohort-level comparisons without requiring routine storage of full segmentation masks. For triage, the output functions as a prioritization aid; it can change reading order and allocate a brief verification step to a subset of studies, while all examinations still undergo full diagnostic interpretation. Patient history and non-cartilage pathology remain part of routine practice and are not replaced by the triage score. Extending triage beyond cartilage and bone would require adding additional structures and broader pathology indicators; our modular design enables these extensions but we did not evaluate them here.

For the longitudinal analysis, the validated biomarkers that enable immediate triage decisions also power our second application. These landmark survival models leverage serial cartilage and meniscus thickness trajectories to predict patient outcomes in the Osteoarthritis Initiative. Once the validity of the decision-relevant biomarkers is confirmed, the pipeline can support both near-term routing and long-term prognostication. Timing results (Supplementary Materials 5.9) support suitability for high volume use. Risk predictions are intended to support care planning and follow-up scheduling rather than to dictate treatment; sites can select operating thresholds using decision-curve analysis (Supplementary Information 5.10) and interpret outputs alongside clinician judgment and patient preferences. Prediction performance may improve by incorporating additional non-imaging covariates when available (e.g., pain scores, prior injury history, comorbidities); our current models use MRI biomarkers plus demographics to isolate the contribution of imaging-derived measurements.

Portability and reuse were central design principles. A single YAML configuration specifies detector type, model weights, classes, and evaluation settings. The system produces standardized numeric biomarkers and summary tables, which helps minimize storage requirements and could enable cross-site comparisons and downstream integration. We will release the code and analysis notebooks and deposit the fine-tuned segmentation weights used in this study on figshare with versioned metadata describing training datasets, label maps, and inference settings.

We acknowledge several limitations. Except for OAI, our datasets were acquired within one health system, and external validation at additional institutions remains needed, especially for hip, shoulder, thigh, and spine imaging. The present evaluation was not designed to stress-test severe motion, substantial susceptibility from hardware, or post-operative implants; performance under these conditions should be evaluated before clinical use. Because several cohorts were curated research or quality-screened clinical datasets, segmentation and biomarker agreement may degrade on less refined real-world scans (e.g., lower SNR, severe motion, metal susceptibility, anisotropic resolution, or uncommon protocols), particularly for thin or low-contrast structures such as cartilage. In those settings, the most reliable strategy will likely combine targeted fine-tuning using representative local scans with conservative automation (QC overlays, range checks, and failure detection), alongside periodic re-validation under protocol drift.

Repeat reads and multi-reader segmentations were not available for every dataset, so we did not compute new inter- or intra-reader variability analyses in this manuscript. Instead, we rely on published annotation protocols and quality checks from the source studies and on biomarker agreement between model-derived and expert-derived measurements as an end-to-end validation signal. Our model comparison was limited to promptable foundation segmenters in the SAM family, and benchmarking alternative architectures, including fully 3D segmentation models that enforce inter-slice continuity, remains future work. Automation relies on detection-generated bounding boxes, which may fail when anatomy is outside the detector’s training distribution or when fields of view differ from those seen during training; in such cases, manual correction of boxes or detector re-training would be required.

Scanner upgrades, reconstruction software changes, and protocol drift can shift image appearance and affect segmentation and downstream biomarkers. A practical deployment would monitor performance on a rolling sample of new scans and apply automated range checks on biomarker outputs. When shifts are detected, re-tuning the segmentation backbone and re-validating biomarker agreement would follow. Although segmentation and biomarker extraction generalize across anatomies, downstream decision thresholds are application-specific; extending decision support beyond the knee will require outcome-linked cohorts and local calibration for each anatomy.

We observed failure modes such as boundary leakage in low-contrast regions and partial under-segmentation in cases with marked tissue loss; errors triggered by prompt shifts also occurred. These cases motivate QC checks such as saving overlay thumbnails, flagging unusual biomarker ranges, and routing low-confidence outputs to human review. We also report lower-tail (5th percentile and minimum) subject-level Dice to make worst-case behavior explicit (Supplementary Data D55–D57). Some measures (meniscal thickness and bone volume) served as proxies to extend tissue coverage and stress-test scalability. Their direct clinical relevance falls outside the present scope and requires dedicated validation. Other prompt strategies, including recently described multimodal text-prompt-driven approaches \cite{zhao_foundation_2025}, may offer different trade-offs in real-time use. Finally, clinician-in-the-loop testing and prospective deployment studies will be needed to determine adoption and operational impact. Any clinical pipelines derived from such segmentation models will need iterative refinement.

This work shows that prioritizing clinical context and system architecture over pure segmentation performance helps shift foundation models from a technology with \enquote{future potential} to one that can drive practical clinical value. These emerging technologies will influence precision medicine, yet the rapid pace of model development can pull attention away from implementation challenges that determine patient impact. A stable measurement layer allows such systems to deliver immediate workflow value (e.g., prioritization and structured reporting) while gradually accumulating standardized longitudinal measurements that future predictive and comparative tools can build on. Our study argues for investment in modular, measurement focused pipelines that absorb technical advances without full architectural redesign; this design choice helps bridge the gap between research innovation and clinical integration. We decouple biomarker extraction from specific model implementations, which creates stable infrastructure for the iterative testing and deployment cycles that real-world adoption requires. Our findings indicate that the path from research to clinical practice depends less on maximizing model performance and more on establishing expert-aligned systems that yield consistent, actionable insights for patient management.


\FloatBarrier


\begingroup

\metaheading{Data availability}
\vspace*{-0.5em} 

All Supplementary Tables (S0–S25) and study-generated Data Tables (D1–D57) cited in the manuscript are publicly available at Figshare: \url{https://doi.org/10.6084/m9.figshare.29633207}.  
Osteoarthritis Initiative (OAI) MRI scans can be accessed via the OAI data portal with registration and data use agreement. Additional institution-specific MRI datasets are subject to institutional review board restrictions; deidentified versions are available from the corresponding author upon reasonable request.  

Fine-tuned segmentation weights used in this study will be deposited in a public repository at the time of publication. Versioned metadata describing training datasets, label maps, and inference settings will be included to support reproduction and external validation.

\vspace*{-0.5em} 
\metaheading{Code availability}
\vspace*{-0.5em} 
All code, configuration files, and preprocessing scripts are available at: \url{https://github.com/gabbieHoyer/AutoMedLabel}. Documentation and environment files are provided for reproducibility. Additional details are described in the Supplementary Engineering Framework.

\vspace*{-0.5em} 
\metaheading{Acknowledgements}
\vspace*{-0.5em} 
This research was supported by the National Institute of Arthritis and Musculoskeletal and Skin Diseases (NIH-NIAMS) through grants R01AR069006, UH3AR076724, R00AR070902, R01AR078762, P50AR060752, R61AR073552, R33AR073552, R01AR0796471, R01AR046905, and R01AR078917. Data and resources from the Osteoarthritis Initiative (OAI) were used in this study. The OAI is a public-private partnership supported by NIH contracts N01-AR-2-2258 through N01-AR-2-2262 and the Foundation for the National Institutes of Health, with contributions from Merck, Novartis, GlaxoSmithKline, and Pfizer. The funders had no role in study design, data acquisition, analysis, interpretation, or manuscript preparation. We thank members of the Musculoskeletal Quantitative Imaging Research (MQIR) group at UCSF for their input and support throughout the development of this work.

\vspace*{-0.5em} 
\metaheading{Author contributions}
\vspace*{-0.5em} 
All authors contributed to the conception and design of the study, as well as to the preparation and approval of the manuscript and Supplementary Materials. G.H. developed the software infrastructure, including metadata management, model fine-tuning, evaluation pipelines, and the automated AutoLabel inference system. Data aggregation and preprocessing were carried out by G.H. and M.W.T., providing a unified basis for analysis. Model training, fine-tuning, and evaluation were conducted by G.H. Statistical design was a collaborative effort among G.H., V.P., and S.M., with G.H. conducting the analysis and V.P. and S.M. performing technical verification. Biomarker evaluation was jointly ideated by all authors, with implementation and analysis executed by G.H.; contributions from M.W.T. and R.B. in data preparation and results validation were integral to the process. G.H. conceptualized and carried out Clinical utility proof-of-concept analyses; validated by S.M. V.P. and S.M. provided leadership in conceptualizing the study, securing funding, and offering valuable input during the manuscript revision process.

\vspace*{-0.5em} 
\metaheading{Competing interests}
\vspace*{-0.5em} 
The authors declare no competing financial or non-financial interests.

\endgroup
\FloatBarrier



\bibliographystyle{sn-nature} 

\FloatBarrier
\clearpage


\begin{shrinkPageWithFooter}[top=1.0cm,bottom=2.0cm,left=2.5cm,right=2.5cm]

\unnumberedsection{Supplementary Tables and Figures}
\unnumberedsubsection{MRI Dataset Overview and Imaging Specs.}

\Needspace{0.9\textheight}
\begin{sideways}
  \begin{minipage}{0.92\textheight}
  
      \makeatletter
        \IfFileExists{images/S0.png}{%
          \makebox[\linewidth][c]{%
            \includegraphics[width=0.87\linewidth]{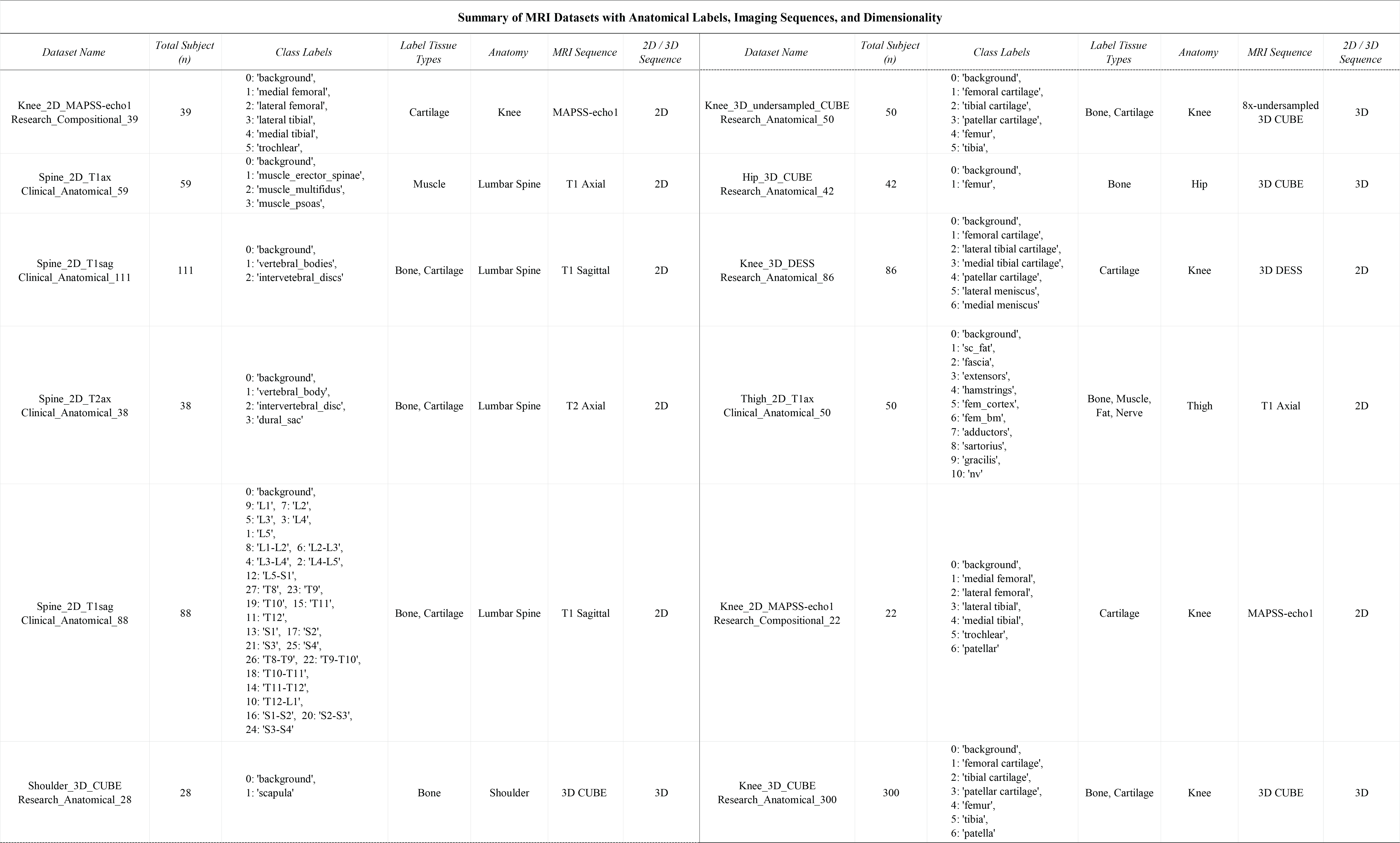}%
          }%
        }{%
          \makebox[\linewidth][c]{%
            \includegraphics[width=0.87\linewidth]{S0.png}%
          }%
        }
        \makeatother

    \captionsetup{type=figure}
    \suppcaptionalm[width=.87\textheight]{%
      \textbf{Table S0: Summary of MRI Datasets with Anatomical Labels, Imaging Sequences, and Dimensionality.} This table summarizes the MRI datasets used in the study, which may serve as a reference for the dataset names, number of subjects, anatomical labels and types, MRI sequences, and dimensionality (2D or 3D). For each dataset, class labels corresponding to anatomical structures such as cartilage, muscles, and bones are included to ensure clarity in segmentation. The datasets are organized by imaged anatomy, MRI sequence type, and dataset size. The naming convention follows the structure \enquote{Anatomy, 2D/3D, MRI sequence, Clinical/Research, Anatomical/Compositional, Dataset Sample Size} to offer a consistent reference point throughout the study.
    }
    \label{fig:supp-s0}
  \end{minipage}
\end{sideways}


\FloatBarrier
\clearpage


\unnumberedsubsection{Fine-tuning Experiments and Evaluation}
\Needspace{0.95\textheight}
\begin{sideways}
  \begin{minipage}{0.95\textheight}
  
    \centering
    \makeatletter
    \IfFileExists{images/S6.png}{%
      \makebox[\linewidth][c]{%
        \includegraphics[width=0.95\linewidth]{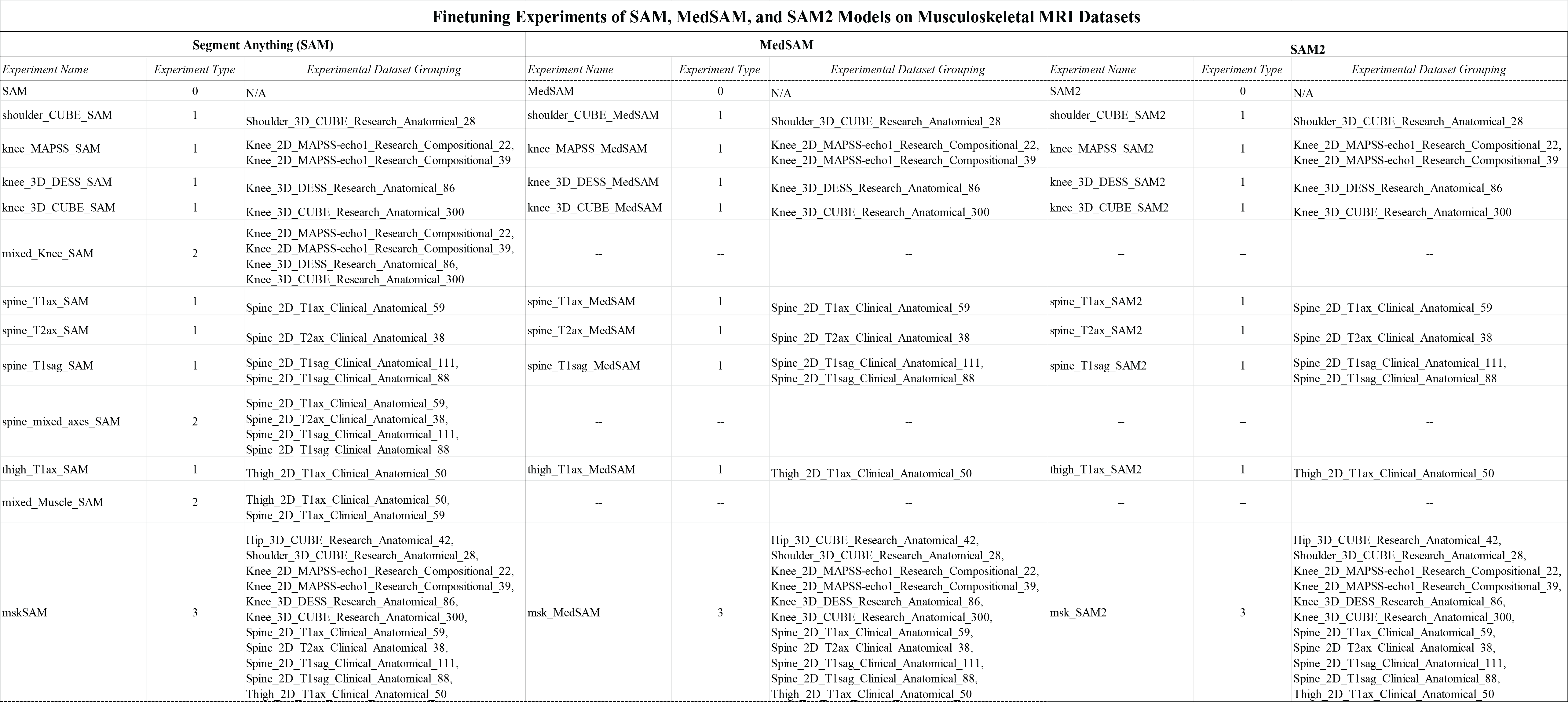}%
      }%
    }{%
      \makebox[\linewidth][c]{%
        \includegraphics[width=0.95\linewidth]{S6.png}%
      }%
    }
    \makeatother

    \captionsetup{type=figure}
    \suppcaptionalm[width=0.9\textheight]{\textbf{Table S6: Fine-tuning Experiments of SAM, MedSAM, and SAM2 Models on Musculoskeletal MRI Datasets.}\protect\\ This table summarizes the fine-tuning experiments conducted with the SAM, MedSAM, and SAM2 models across various musculoskeletal (MSK) MRI datasets. Each experiment lists the name, type, and the datasets used for fine-tuning. The experiments encompass anatomical regions such as the knee, shoulder, spine, hip, and thigh, with models fine-tuned on individual datasets or combinations thereof (e.g., mixed or fused datasets). The experiment type indicates whether training involved a single dataset or multiple datasets. This overview of fine-tuning strategies facilitates the evaluation of model adaptability and performance across diverse MSK MRI imaging conditions and dataset configurations.}

    \label{fig:supp-s6}
  \end{minipage}
\end{sideways}

\FloatBarrier


\bgroup
\fixFloatSize{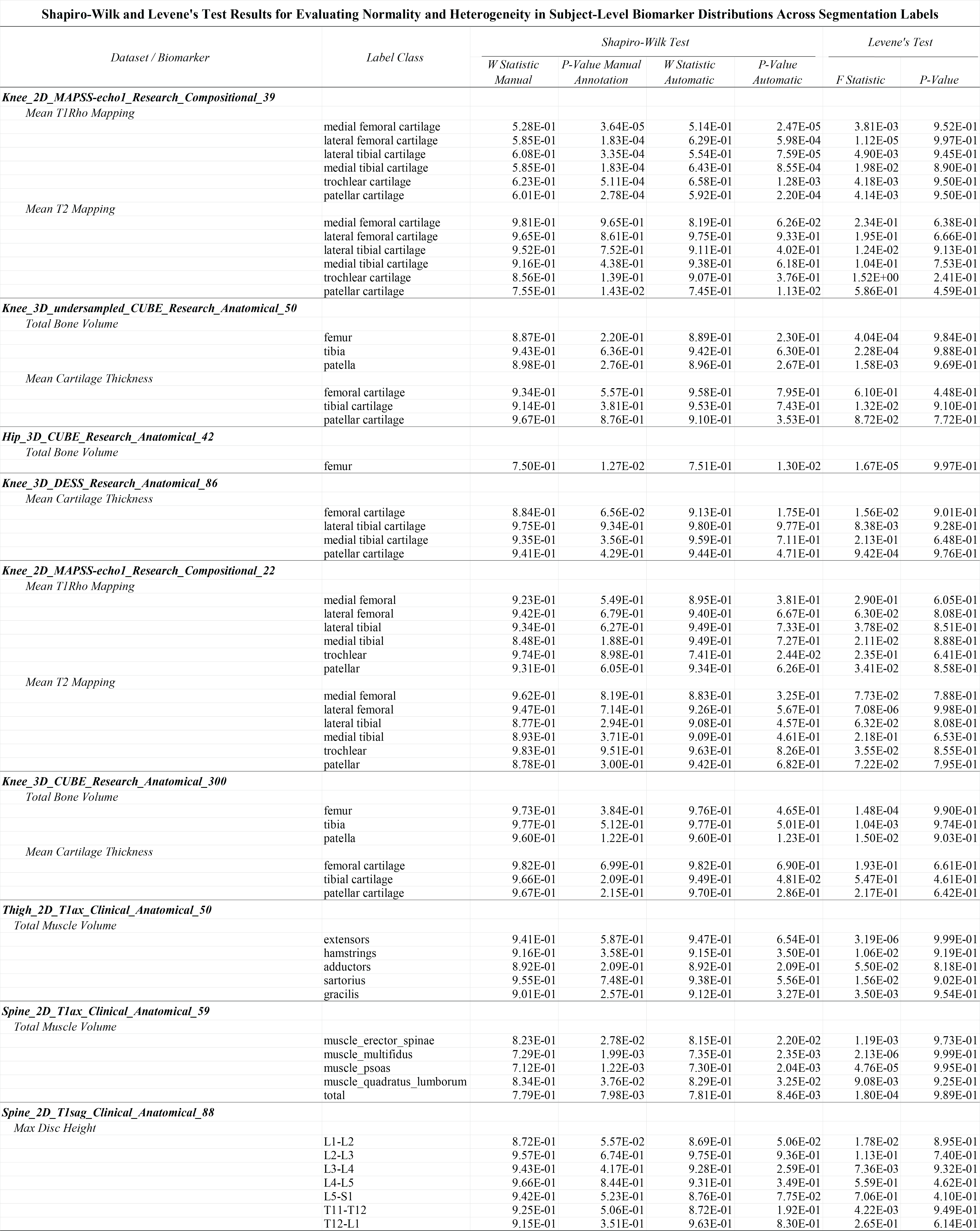}
\begin{figure*}[!htbp]

\unnumberedsubsection{Biomarker Metrics and Evaluation}

\vspace{3em}
\phantomsection
\label{fig:supp-s13}

\makeatletter
\IfFileExists{images/S13.png}{%
  \makebox[\linewidth][c]{%
    \includegraphics[width=0.9\linewidth]{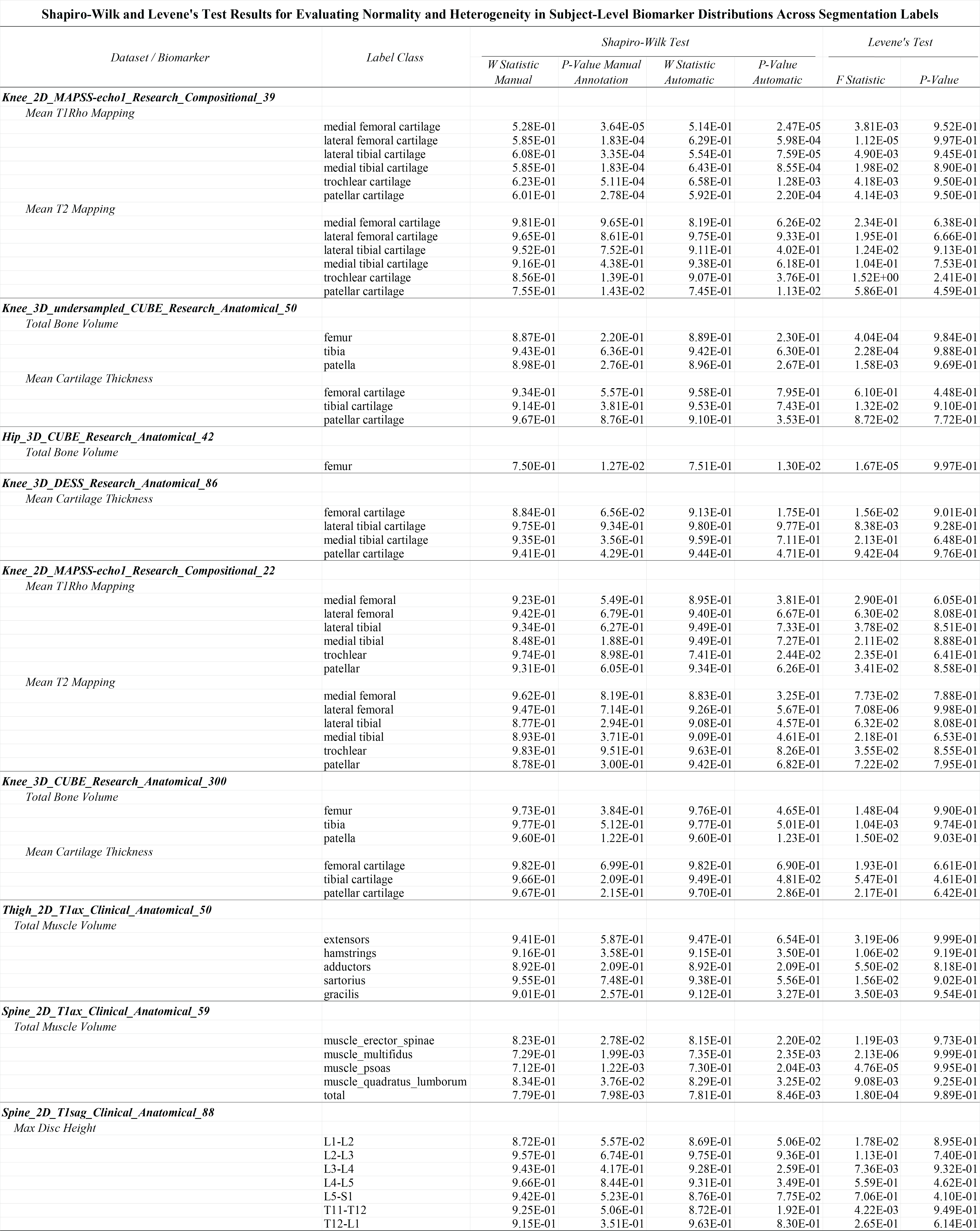}%
  }%
}{%
  \makebox[\linewidth][c]{%
    \includegraphics[width=0.9\linewidth]{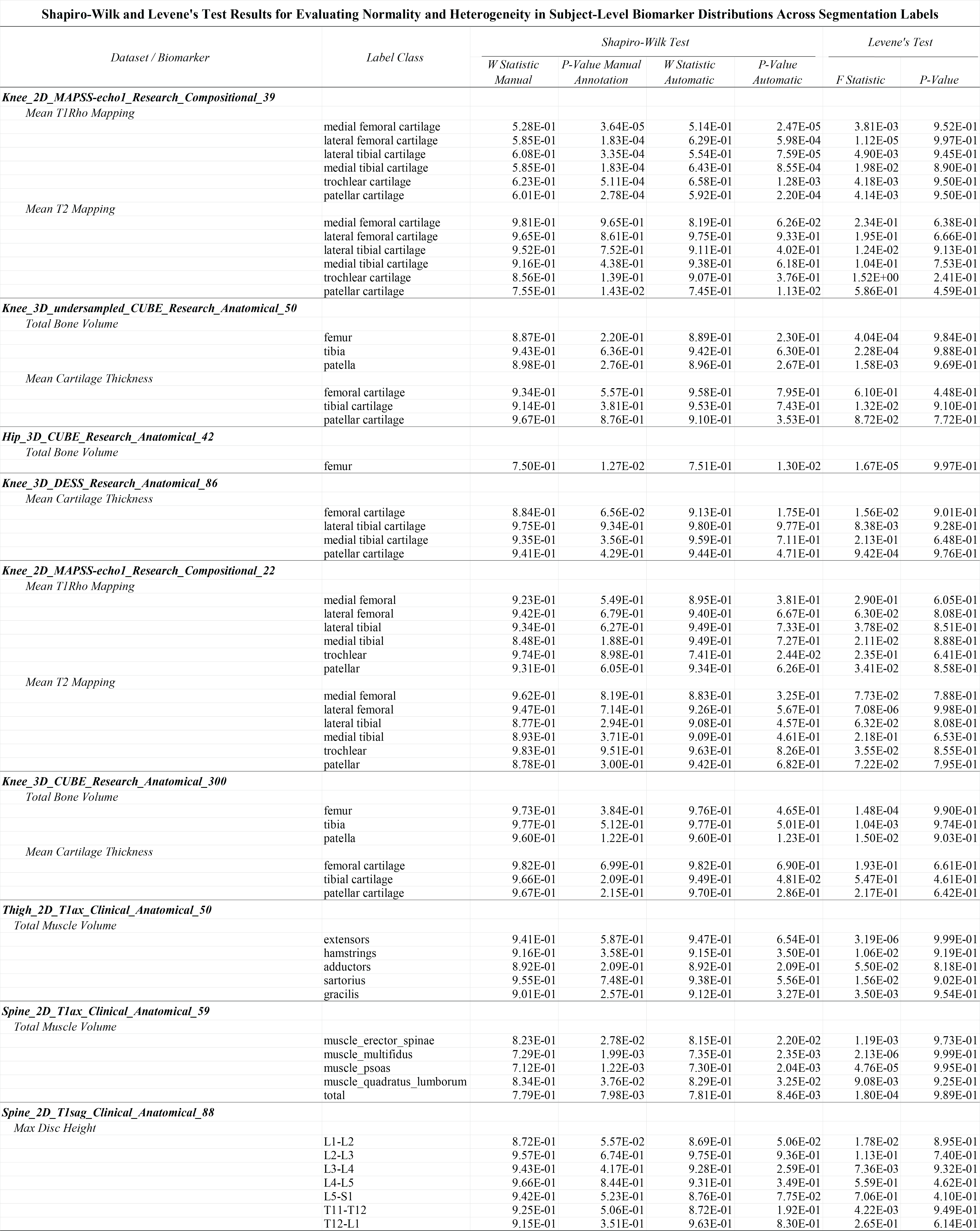}%
  }%
}
\makeatother

\begin{minipage}{\linewidth}
\small\justify
\suppcaptiontiny[width=1.0\linewidth]{\textbf{Table S13: Shapiro-Wilk and Levene's Test Results for Evaluating Normality and Homogeneity in Subject-Level Biomarker Distributions Across Segmentation Labels.} This table presents the results of the Shapiro-Wilk test for assessing normality and the Levene test for evaluating homogeneity of variances in subject-level biomarker distributions across various segmentation labels. These biomarkers are derived from both manual annotations and automatic model inferences across datasets. Metrics evaluated include mean cartilage thickness, total bone volume, total muscle volume, and disc height, among others. The Shapiro-Wilk test, a two-sided test for normality, was applied to both manually annotated and automatically generated biomarker distributions to evaluate whether the data follow a normal distribution. For each biomarker distribution, the table reports the Shapiro-Wilk W~statistic and the associated p-value for both methods. The W~statistic measures how well the data conform to a normal distribution, and smaller p-values indicate significant deviations from normality, suggesting non-parametric methods should be used if p \textless\ 0.05. Levene's test, also two-sided, was used to assess the homogeneity of variances between the biomarker distributions from manual and automatic segmentations. The median was used as the center to compare variances, making the test robust for skewed distributions. For each segmentation label, the table reports the Levene F statistic and the associated p-value. A p \textless\ 0.05 indicates a significant difference of the variances between the two methods differ significantly, suggesting that homogeneity assumptions for parametric testing may not hold. These tests guide the determination of whether parametric or non-parametric methods should be applied in subsequent analyses based on whether normality and homogeneity assumptions are met.}
\end{minipage}

\end{figure*}
\egroup

\FloatBarrier
\clearpage


\bgroup
\fixFloatSize{images/ad76d1d5-349c-4430-8e28-4f339efd6fc1-us14.png}
\begin{figure*}[!htbp]
\phantomsection
\label{fig:supp-s14}

\makeatletter
\IfFileExists{images/S14.png}{%
  \makebox[\linewidth][c]{%
    \includegraphics[width=1.1\linewidth]{images/S14.png}%
  }%
}{%
  \makebox[\linewidth][c]{%
    \includegraphics[width=1.1\linewidth]{S14.png}%
  }%
}
\makeatother

\suppcaptionmed[width=1.1\linewidth]{\textbf{Table S14: Intraclass Correlation Coefficient (ICC) Analysis for Biomarker Metrics Comparing Manual and Automated Segmentation Across MRI Datasets.} This table presents the results of the Intraclass Correlation Coefficient (ICC) analysis, evaluating the consistency of biomarker metrics between manual annotations and automated segmentation methods across MRI datasets. Two ICC approaches are displayed: a parametric ICC3 model for biomarker distributions meeting normality and homogeneity assumptions, and a non-parametric ICC derived from a linear mixed-effects model with bootstrap resampling (N = 10,000) for distributions failing those assumptions. \mbox{}\protect\newline\newline The table includes ICC values, F-statistics, degrees of freedom (df), p-values, and 95\% confidence intervals (CI) for both approaches. While the parametric ICC3 results are included for clarity and comparison, the non-parametric ICC is emphasized as it better captures complex patterns and provides a more accurate representation of agreement when distributional assumptions are not met. The mixed model incorporates subject-level variability by treating subjects as random effects, producing ICC values based on variance components. \mbox{}\protect\newline\newline Orange-highlighted cells indicate cases where the linear ICC was computed for non-normal distributions, showing overly optimistic performance compared to the more conservative and realistic non-parametric ICC values in the adjacent columns. Higher ICC values in the non- parametric results reflect stronger agreement between manual and automated methods; this provides a robust basis for assessing segmentation reliability across varying biomarker metrics.}

\end{figure*}
\egroup
\FloatBarrier

\clearpage 


\bgroup
\fixFloatSize{images/10e5388d-67d3-4eeb-8603-4e96a55fe634-us18.png}
\begin{figure*}[!htbp]
\phantomsection
\label{fig:supp-s18}

\unnumberedsubsection{Clinical Utility Validation}
\vspace{3em}

\makeatletter
\IfFileExists{images/S18.png}{%
  \makebox[\linewidth][c]{%
    \includegraphics[width=1.1\linewidth]{images/S18.png}%
  }%
}{%
  \makebox[\linewidth][c]{%
    \includegraphics[width=1.1\linewidth]{S18.png}%
  }%
}
\makeatother

\begin{minipage}{\linewidth}

\suppcaptionmed[width=1.1\linewidth]{\textbf{Table S18: Performance of the three stage knee MRI triage pipeline.} Stage A screens the full cohort for any abnormality. Knees that pass Stage A enter Stage B, which distinguishes cartilage \ensuremath{+} bone findings from all other cases. Stage C is applied only to knees that pass Stage B and is divided into (i) Part 1, which flags joint-level abnormality (bone \ensuremath{+} cartilage) for femur, tibia and patella, and (ii) Part 2, which provides separate bone- versus cartilage-specific scores for each joint. \mbox{}\protect\newline \ensuremath{\bullet} \hspace*{0.5em}AUC values are calculated once on the complete out-of-fold predictions. \mbox{}\protect\newline \ensuremath{\bullet} \hspace*{0.5em}\enquote{AUC lo / hi} and \enquote{sens lo / hi} are the 2.5th and 97.5th percentiles from 2000 non-parametric bootstrap samples. \mbox{}\protect\newline \ensuremath{\bullet} \hspace*{0.5em}\enquote{sens} is the sensitivity at the fixed specificity shown in spec\%. \mbox{}\protect\newline \ensuremath{\bullet} \hspace*{0.5em}{thr} is the probability threshold that attains the target specificity and is reported only for stages that set routing cut-offs (Stage A and Stage B). Stage C inherits the routed subset, so no new threshold is defined, and the column is marked \enquote{{\textendash}}. \mbox{}\protect\newline \ensuremath{\bullet} \hspace*{0.5em}Operating-point labels: \mbox{}\protect\newline \hspace*{1em}\ensuremath{\circ} \hspace*{0.5em}all {\textendash} Stage A evaluated at 90\% specificity (screening step)  \mbox{}\protect\newline \hspace*{1em}\ensuremath{\circ} \hspace*{0.5em}B85 / B90 - Stage B evaluated at 85\% or 90\% specificity, respectively \mbox{}\protect\newline\newline This format allows direct comparison of point estimates with their 95\% confidence intervals at the operating points used in the triage workflow.
}
\end{minipage}

\end{figure*}
\egroup
\FloatBarrier

\clearpage 


\bgroup
\fixFloatSize{images/S_Fig17.png}

\begin{sidewaysfigure*}
\phantomsection
\label{fig:supp-fig-s17}

\centering 
\def\tsGraphicsScaleY{.7}%

\makeatletter\IfFileExists{images/S_Fig17.png}{\includegraphics[width=0.9\linewidth]{images/S_Fig17.png}}{\includegraphics[width=0.9\linewidth]{S_Fig17.png}}

\makeatother 
\suppcaption[width=0.9\linewidth]{\textbf{Supplementary Figure 17: ROC and survival curves.} Panels a–d show ROC curves for five algorithms (LR, RF, XGB, vote, stack) evaluated at the 48-month landmark; panels e–h show Kaplan-Meier curves that stratify the test set by the RF predicted risk (median split). \mbox{}\protect \textbf{a)} TKR 48–96m \textbf{b)} TKR 48–120m \textbf{c)} OA incidence 48–96m \textbf{d)} OA incidence 48–120m
\mbox{}\protect\newline \textbf{e–h)} Corresponding survival curves demonstrate clear risk separation for TKR horizons, modest separation for OA incidence.
}
\end{sidewaysfigure*}
\egroup
\FloatBarrier
\clearpage 


\bgroup
\fixFloatSize{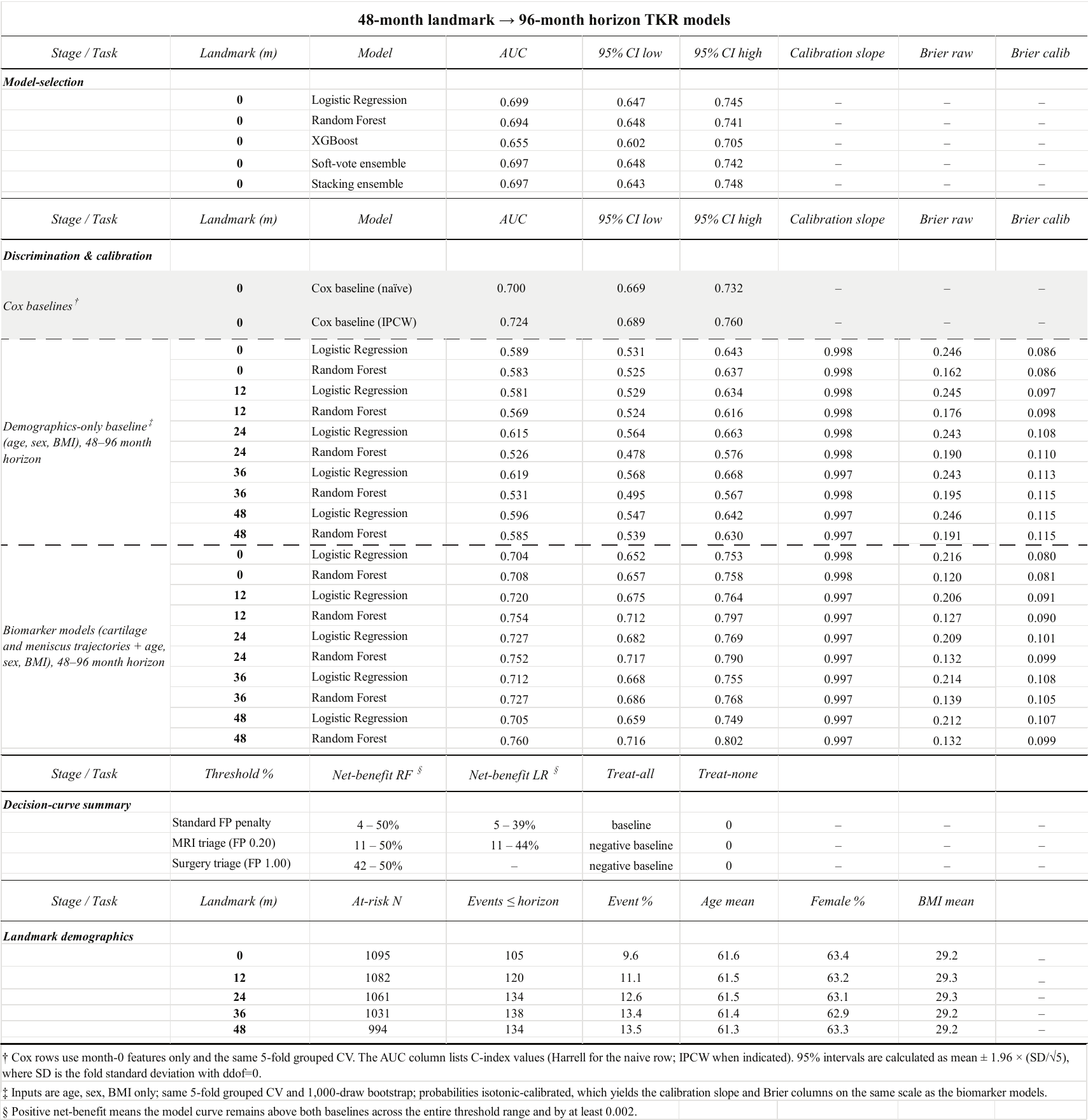}
\begin{figure*}[!htbp]
\phantomsection
\label{fig:supp-s20}

\makeatletter
\IfFileExists{images/S20_baselines.pdf}{%
  \makebox[\linewidth][c]{%
    \includegraphics[width=1.1\linewidth]{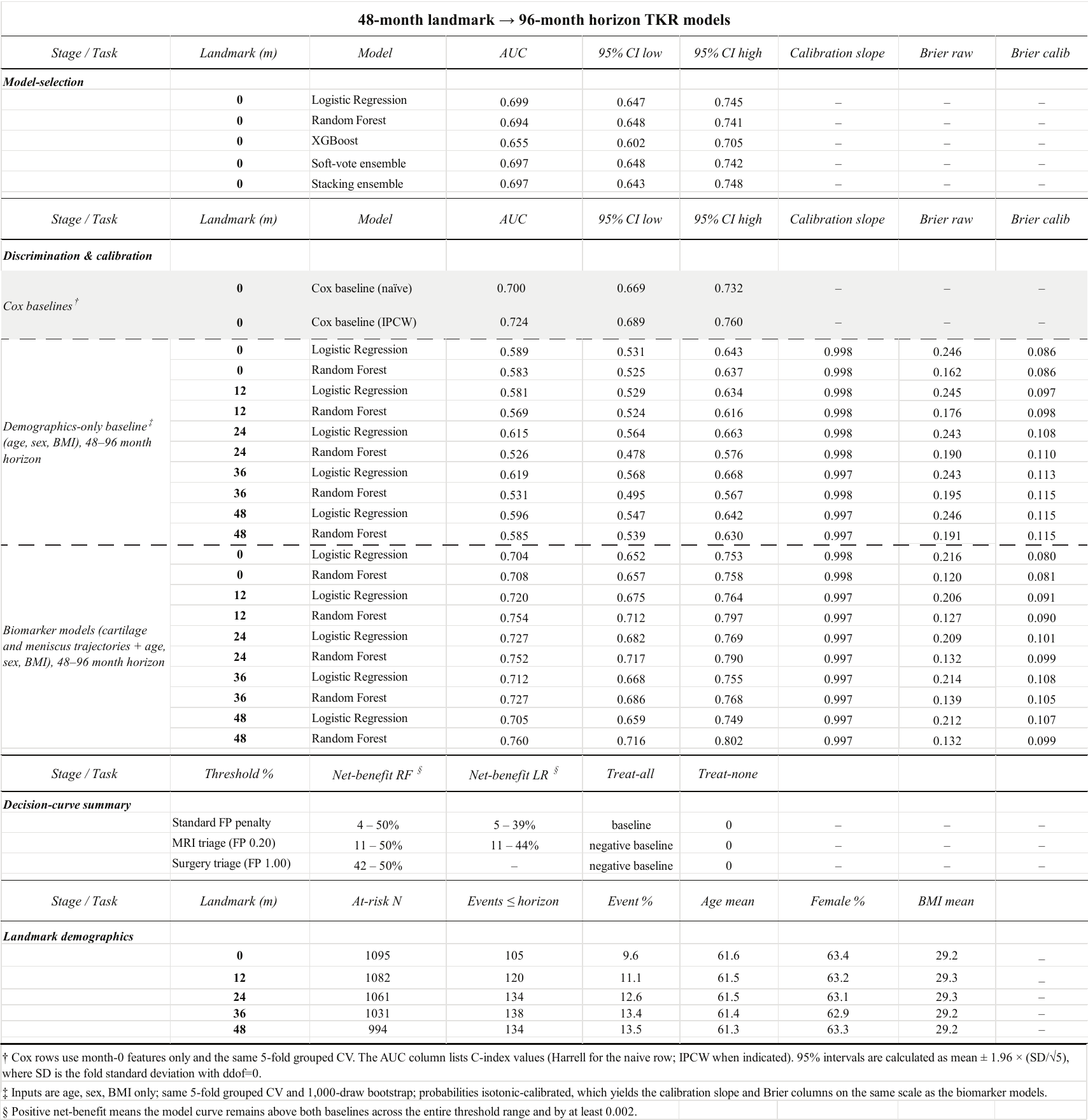}%
  }%
}{%
  \makebox[\linewidth][c]{%
    \includegraphics[width=1.1\linewidth]{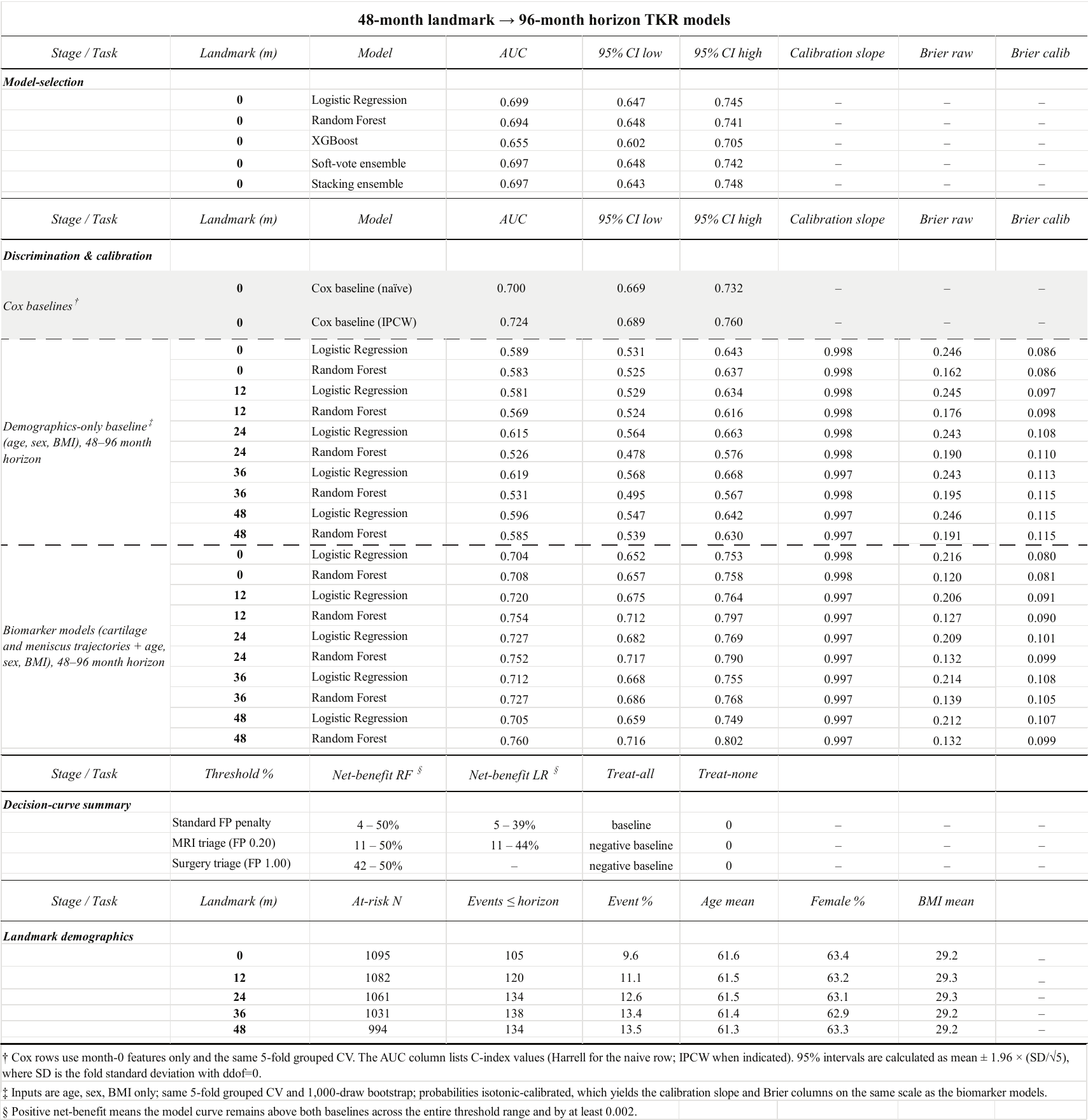}%
  }%
}
\makeatother

\suppcaptionalm[width=1.1\linewidth]{\textbf{Table S20: Performance of knee-replacement risk models trained on biomarkers up to 48 months and tested on the 48-96 month horizon.} Columns report discrimination (AUC or C-index with 95\% confidence intervals), calibration slope, raw and isotonic-calibrated Brier scores, decision-curve net-benefit ranges, and demographic counts for knees still at risk at each landmark. Model-selection AUCs are limited to the 0-month landmark. Cox proportional-hazards baselines use month-0 features only\textsuperscript{†}. Demographics-only baselines (age, sex, BMI) are listed at each landmark\textsuperscript{‡}. Net-benefit windows are shown when the model curve exceeds both \enquote{treat-all} and \enquote{treat-none} by at least 0.002\textsuperscript{§}.}
\end{figure*}
\egroup

\FloatBarrier

\end{shrinkPageWithFooter}


\end{document}